\documentclass[%
 reprint,
 amsmath,amssymb,
 aps,
prd,
showkeys
]{revtex4-1}

\usepackage{graphicx}
\usepackage{dcolumn}
\usepackage{bm}
\usepackage{xcolor}

\begin{document}


\title{An analytical formula for signal optimization in stimulated photon-photon scattering setup with three laser pulses}

\author{A.V.~Berezin}
 \email{arsenbrs@mail.ru}
 \affiliation{%
 National Research Nuclear University ``MEPhI'', 31, Kashirskoe Highway, Moscow, 115409 Russia
}%
 
\author{A.M.~Fedotov}
 \email{al.m.fedotov@gmail.com}
\affiliation{%
 National Research Nuclear University ``MEPhI'', 31, Kashirskoe Highway, Moscow, 115409 Russia
}%

\date{\today}

\begin{abstract}
We consider a setup to detect stimulated photon-photon scattering using high-power lasers. Signal photons are emitted from an overlap of the incoming intense laser pulses focused in vacuum from three sides. We derive and justify a general approximate analytical formula for the angular distribution and total yield of such signal photons in terms of  the parameters of the incoming pulses, including their intensity, carrier frequencies, durations, focusing,  polarizations, mutual orientation and overlap. Using the obtained formula a parametric study of the signal is carried out and  optimization is performed.
\end{abstract}

                              
\maketitle

\section{Introduction}
From the modern perspectives, strong electromagnetic fields modify quantum fluctuations in a vacuum, changing its properties. This is called vacuum polarization and is described by the Heisenberg-Euler effective action \cite{euler1936consequences, berestetskii1982quantum}
\begin{equation}\label{action}
S = \! \int d^4x \left( 
\frac{\mathfrak{F}}{4\pi} + \frac{\alpha}{360 \pi^2 E_c^2} 
(4 \mathfrak{F}^2 + 7 \mathfrak{G}^2) + \dots \right),
\end{equation}
where $\mathfrak{F} = (E^2 - H^2) / 2$ and $\mathfrak{G} = \mathbf{E} \cdot \mathbf{H}$ are the electromagnetic field invariants, $E_c = m^2c^3 \! / (e\hbar) = 1.3 \times 10^{16}$~V/cm is the critical field, $\alpha = e^2 \! / (\hbar c)$ is the fine structure constant. Here $m$ is the electron mass, $e$ is the magnitude of the electron charge, $c$ is the speed of light in vacuum, and $\hbar$ is the reduced Planck constant. Expression \eqref{action} is valid for fields weaker than the critical field $E_c$ and varying slowly over the Compton length and over the Compton time. The terms additional to the action of classical electromagnetism called radiative corrections are of quantum nature. We consider only the leading radiative correction called the four-wave mixing and given explicitly in Eq.~\eqref{action}, since for laser fields under consideration the higher-order terms are about $\propto 10^{-6}$ times smaller.

Radiative corrections in Eq.~\eqref{action} bring additional nonlinear terms to Maxwell equations. In a variable field these terms act as sources for a detectable emission of real photons from polarized vacuum. In quantum language and when considering only the main four-wave interaction corrections, such radiation is associated with the process of elastic photon-photon scattering \cite{euler1935scattering, karplus1950non, karplus1951scattering, de1964dispersive, costantini1971nonlinear, ahmadiniaz2022worldline}. Significant deviations of the coefficients from the values given in Eq.~\eqref{action}, if exceed the accuracy of the supposed approximations, may indicate a new physics, in particular, the contribution of axion-like components of dark matter. Up to date, the effect of real photon-photon scattering has been never observed directlly  \cite{vavilov1928zamechaniya, vavilov1930attempt, hughes1930attempt, moulin1996photon, bernard2000search, battesti2012magnetic, fedotov2022advances}. 

Theoretical analysis of three-pulse experimental schemes for detecting photon-photon scattering was started quite long ago \cite{varfolomeev1966induced, dewar1974induced, grynberg1990possibilite}. The supplement of a third pulse stimulates the scattering process, enhancing the feasibility of the effect in comparison with earlier two-pulse schemes. However, the early works ignored the real structure of the fields of the focused laser pulses, which were represented just as plane monochromatic waves, allowing at best only to estimate the magnitude of the effect. The most interesting development of these works was the invention of more favorable collision geometries, for which the signal photons to be detected were better separated from the initial pulses in direction and frequency \cite{bernard2000search, lundstrom2005detection, lundstrom2006using}. Furthermore, the geometric factors were obtained in Ref.~\cite{lundstrom2005detection, lundstrom2006using} for arbitrary linear polarization of the pulses.

More recent advance of the three-pulse schemes was inspired by the progress of high-power lasers that would make experimental detection of the effect feasible. In particular, the structure of the focused pulsed field was refined by applying the Gaussian beam model \cite{goubau1961guided, siegman13lasers}. At the same time, using more sophisticated field models did not allow to  carry out all calculations analytically, so that the final signal distributions could be calculated only numerically for some particular accepted values of the parameters \cite{king2012photon, gies2018photon, king2018three}. Some authors considered collision schemes involving a larger number of pulses \cite{klar2020detectable, gies2021quantum}. Their advantage is a richer set of potentially detected signals, but the distributions become less clear and more difficult to read off.

Here we advance the consideration of three-pulse schemes for detecting photon–photon scattering by deriving an analytical formula for the total yield of the signal photons. The formula takes into account realistic structure of the laser fields and contains the dependence on the full set of parameters (peak power, duration, focusing width, polarization) of each of the pulses. In addition, it applies to any geometry of the collision setup.

The paper is organized as follows. In Sec.~\ref{sec2} we review the basic calculation formulas and the model of a focused Gaussian beam with an arbitrary polarization. In Sec.~\ref{sec3} the most general form geometry of a three-pulse collision scheme is considered and a universal analytical formula for the signal photons yield is obtained. Thereafter, using the obtained formula, in Sec.~\ref{sec4} the dependence of the signal on the key parameters is obtained for specific example collision scheme and, thereby, the conditions maximizing it are discussed. In Sec.~\ref{sec5} we conclude. The appendices include the discussion of the parameters of a Gaussian pulse in terms of the characteristics of real lasers and the decryption of the (rather complicated) expressions  encountered in the derived analytical formulas.

\section{Formalism} \label{sec2}

\subsection{Number of emitted photons}
As follows from Eq.~\eqref{action}, an external electromagnetic field induces the vacuum sources 
\begin{equation} \label{sources}
\rho = -\text{div} \, \mathbf{P}, \qquad
\mathbf{j} = \frac1{c}\frac{\partial \mathbf{P}}{\partial t} - \text{rot} \, \mathbf{M},
\end{equation}
where the polarization $\mathbf{P}$ and magnetization $\mathbf{M}$ of the vacuum in the same approximation as Eq.~\eqref{action} are determined by the expressions
\begin{eqnarray}
\notag
\mathbf{P} &=& \frac{\alpha}{180 \pi^2 E_c^2} 
(4 \mathfrak{F} \mathbf{E} + 7 \mathfrak{G} \mathbf{H}), 
\\
\mathbf{M} &=& \frac{\alpha}{180 \pi^2 E_c^2} 
(4 \mathfrak{F} \mathbf{H} - 7 \mathfrak{G} \mathbf{E}).
\end{eqnarray}

In the considered case of stimulated scattering, following \cite{Galtsov:1971xm,fedotov2007generation,karbstein2015stimulated,gies2018all},  the average number of signal photons emitted in a narrow frequency range $(\omega, \, \omega + d\omega)$ can be found by dividing the energy emitted by sources \eqref{sources} in this spectral interval, calculated according to classical electrodynamics \cite{landau2013course},
by the photon energy $\hbar\omega$. Hence, for the average total number of emitted signal photons, we find
\begin{eqnarray} \label{N}
N &=& \int \frac{d \Omega_\mathbf{\hat{k}}}{4\pi}
\int\limits_{0}^{+\infty} \frac{\omega^3 d\omega}{\pi \hbar c^3}
\nonumber
\\
& &\times \left| \int d^4 x \left( \mathbf{\hat{k}} \times \mathbf{P} +
\mathbf{\hat{k}} \times \left( \mathbf{\hat{k}} \times \mathbf{M} \right) \right)
e^{ikx} \right|^2.
\end{eqnarray}
Here and below, we use a standard abbreviation for the scalar product in Minkowski space
\begin{equation}
k x \equiv k^\mu x_\mu = \omega 
\left(t - \mathbf{\hat{k}} \cdot \mathbf{r}/c \right), \qquad
k^\mu = \omega (1, \mathbf{\hat{k}}) / c.
\end{equation}
The integrand in Eq.~\eqref{N} determines the average spectrum and angular distribution of the signal.
Since signal photons are emitted in a coherent state, the uncertainty in their actual number is of the order of $\Delta N\simeq \sqrt{N}$.

\subsection{Focused laser pulse model}

In what follows we consider the emission of signal photons from an overlap of a number of colliding focused laser pulses.
The field of the $l$-th pulse propagating along the direction $\mathbf{\hat{k}}_l$,
can be represented as
\begin{equation} \label{ELHL}
\mathbf{E}^{L_l} = \textrm{Re}\, \left( \mathbf{E}^{L_l}_s e^{-ik_lx} \right), \quad
\mathbf{H}^{L_l} = \textrm{Re}\, \left( \mathbf{H}^{L_l}_s e^{-ik_lx} \right),
\end{equation}
where $\omega_{l}$ is the carrier frequency of the pulse, and
$\mathbf{E}^{L_l}_s(t, \mathbf{r})$, $\mathbf{H}^{L_l}_s(t, \mathbf{r})$ are complex vector envelopes, slowly varying compared to the oscillating factor. For simplicity, all the pulses are considered Gaussian \cite{goubau1961guided, siegman13lasers}, so that
for a pulse propagating along the axis $z$ ($\mathbf{\hat{k}}_l = \mathbf{\hat{e}}_z$) and passing the center of the focus at the origin at $t=0$, the corresponding complex vectors $\tilde{\mathbf{E}}^{L_l}_s(t, \mathbf{r})$, $\tilde{\mathbf{H}}^{L_l}_s(t, \mathbf{r})$ have the form
\begin{equation} \label{beaml}
\tilde{\mathbf{E}}^{L_l}_s =  
\mbox{\boldmath $\epsilon$}_l A_l 
e^{i\varphi_{0, \, l}}, \qquad
\tilde{\mathbf{H}}^{L_l}_s = \mathbf{\hat{e}}_z \times \tilde{\mathbf{E}}^{L_l}_s, 
\end{equation}
where $\mbox{\boldmath $\epsilon$}_l = \left( \cos \theta_l, \, e^{i \delta_l} \sin \theta_l, \, 0 \right)$ is the normalized complex polarization vector \cite{jekrard1954transmission}, $\varphi_{0,\,l}$ is the carrier-envelope phase,
\begin{equation} \label{Al}
A_l(t, \mathbf{r}) = \frac{A_{0, \, l}}{\kappa_{l}}
\exp \left( -\phi_{\parallel, \, l}^2 - \frac{\phi_{\perp, \, l}^2}{\kappa_{l}}
\right), 
\end{equation}
and
\begin{eqnarray} \label{alDeltal}
\notag
\phi_{\parallel, \, l} = a_l \omega_{l} (z/c-t), &\quad&
\phi_{\perp, \, l} = \Delta_l \omega_{l} \sqrt{x^2 + y^2} / c, 
\\
\kappa_{l} = 1 + i \phi_{z, \, l}, &\quad&
\phi_{z, \, l} = 2 \Delta_l^2 \omega_{l} z / c. 
\end{eqnarray} 

In Eqs.~\eqref{Al}, \eqref{alDeltal} the amplitude $A_{0, \, l}$ is determined by the peak power $P_l$ of the pulse, and the dimensionless small parameters $a_l\ll1$ and $\Delta_l\ll1$ characterize the duration $\tau_l$ of the pulse and its focal width $w_l$ in comparison with the wave period and wavelength $\lambda_l = 2 \pi c / \omega_{l}$, respectively (see Appendix \ref{app1} for details).

The expressions for envelopes of an arbitrarily directed pulse  are constructed by appropriate shift and rotation:
\begin{eqnarray}
\notag
\mathbf{E}^{L_l}_s (t, \mathbf{r}) &=& 
M_l \cdot \tilde{\mathbf{E}}^{L_l}_s 
\left(t-t_l, M_l^{-1} \cdot (\mathbf{r}-\mathbf{r}_l) \right), 
\\
\mathbf{H}^{L_l}_s (t, \mathbf{r}) &=& 
M_l \cdot \tilde{\mathbf{H}}^{L_l}_s 
\left(t-t_l, M_l^{-1} \cdot (\mathbf{r}-\mathbf{r}_l) \right),
\end{eqnarray}
where the rotation matrices $M_l$ in the selected coordinate system $(\mathbf{\hat{e}}_x, \, \mathbf{\hat{e}}_y, \, \mathbf{\hat{e}}_z)$ are determined by the actual directions of the pulses:
\begin{equation}
M_l (\mathbf{\hat{k}}_l) =  \left( \frac{\mathbf{\hat{e}}_z \times \mathbf{\hat{k}}_l}
{|\mathbf{\hat{e}}_z \times \mathbf{\hat{k}}_l|}, \, 
\frac{\mathbf{\hat{k}}_l \times (\mathbf{\hat{e}}_z \times \mathbf{\hat{k}}_l)}
{|\mathbf{\hat{e}}_z \times \mathbf{\hat{k}}_l|}, \, \mathbf{\hat{k}}_l \right)^{-1}.
\end{equation}

\section{Analytical formula for the number of signal photons} \label{sec3}

Eq.~\eqref{N} is our starting one. The internal (spatio\,--\,temporal) integral is four-dimensional within infinite limits, in case of moderate focusing and long pulse duration (as compared to the wavelength and period, respectively) with a rapidly oscillating integrand. Thus its direct numerical evaluation by standard methods repeated for different values of parameters is challenging in terms of both the required computational resources and time. However, the method described below allows not only to circumvent this difficulty by significantly simplifying and optimizing calculations, but also to derive an approximate analytical formula applicable in a wide range of parameters.

\subsection{Isolation of fundamental harmonics}
We assume that laser pulses~$\{ L_l \}$ collide coherently, so that the electric and magnetic fields in the region of their overlap are coherent superpositions of their individual fields,
\begin{equation} \label{EH}
\mathbf{E} = \sum\limits_{l} \mathbf{E}^{L_l}(t, \mathbf{r}), \qquad
\mathbf{H} = \sum\limits_{l} \mathbf{H}^{L_l}(t, \mathbf{r}).
\end{equation}

Consider separately the integral $\int d^4x\, \mathbf{P} e^{ikx}$ contained in Eq.~\eqref{N} (the below consideration holds also for the integral of magnetization $\int d^4x\, \mathbf{M} e^{ikx}$, hence for the whole integrand in Eq.~\eqref{N}). Taking into account Eqs.~\eqref{ELHL} and~\eqref{EH}, the integrand is represented as a sum of harmonics oscillating at constant frequencies:
\begin{equation} \label{intP}
\int d^4x\, \mathbf{P} e^{ikx} = \frac{\alpha}{180 \pi^2 E_c^2} 
\int d^4x \!\!\! \sum_{l_1, \, l_2, \, l_3} \!\!\! \mathbf{G}_{l_1 l_2 l_3}, 
\end{equation}
where the factor $\mathbf{G}_{l_1 l_2 l_3}$ contains a product of the complexified fields of the pulses $L_{l_1}$, $L_{l_2}$ and $L_{l_3}$ or their complex conjugates. For example,
\begin{equation} \label{harmonic}
\mathbf{G}_{12}^3 = \mathbf{p}_{12}^{3} A_1 A_2 A_3^*
e^{-i (k_1 + k_2 - k_3 - k) x},
\end{equation}
where
\begin{equation} \label{p123}
\mathbf{p}_{12}^3 \equiv 
e^{i (k_1 + k_2 - k_3)x}
\frac{\partial^3}{\partial A_1 \partial A_2 \partial A_3^*} 
\left( 4 \mathfrak{F} \mathbf{E} + 7 \mathfrak{G} \mathbf{H} \right)
\end{equation}
is a constant vector determined by the propagation directions and polarizations of the pulses (see Appendix \ref{app2}), and the upper index ``3'' means that the field of the third pulse is taken complex conjugate. 

Contributions of such harmonics to the integral in Eq.~\eqref{intP} is suppressed due to their fast oscillation in space and time, unless the argument of the overall rapidly oscillating exponential factor vanishes. This corresponds to fulfillment of the energy\,--\,momentum conservation law. Hence the values of the frequency $\omega$ and propagation direction $\mathbf{\hat{k}}$ that provide such an equality determine the maxima of the spectral and angular distributions of the emitted photons.

Let us classify the harmonics by their type. Single-pulse harmonics $\mathbf{G}_{l l l}$, $\mathbf{G}_{ll}^l$, $\mathbf{G}_{l}^{ll}$, $\mathbf{G}^{lll}$ are absent in the expansion \eqref{intP} due to the vanishing of the field invariants~$\mathfrak{F}$,~$\mathfrak{G}$ for the field of each pulse \eqref{ELHL} in our approximation. Two-pulse harmonics $\mathbf{G}_{l_1 l_1 l_2}$, $\mathbf{G}_{l_1 l_1}^{l_2}$, $\mathbf{G}_{l_1 l_2}^{l_1}$, $\mathbf{G}_{l_1}^{l_1 l_2}$, $\mathbf{G}_{l_2}^{l_1 l_1}$, $\mathbf{G}^{l_1 l_1 l_2}$ + $l_1 \leftrightarrow l_2$ also contribute to \eqref{intP}, but the argument of their oscillating factor either cannot vanish at all, or the maxima of the distributions of the radiated photons coincide in direction and frequency with either one of the initial pulses, thus making their experimental detection hardly possible. For this reason, three-pulse harmonics $\mathbf{G}_{l_1 l_2 l_3}$, $\mathbf{G}_{l_1}^{l_2 l_3}$, $\mathbf{G}_{l_2}^{l_1 l_3}$, $\mathbf{G}_{l_3}^{l_1 l_2}$, $\mathbf{G}^{l_1 l_2 l_3}$, $\mathbf{G}_{l_2 l_3}^{l_1}$, $\mathbf{G}_{l_1 l_3}^{l_2}$, $\mathbf{G}_{l_1 l_2}^{l_3}$ are of most interest, moreover only last three of them are such that the argument of the oscillating factor can ever vanish (as the remaining transitions are forbidden by conservation laws). Therefore, as most of other authors  \cite{varfolomeev1966induced, dewar1974induced, bernard2000search, lundstrom2006using, king2012photon, tennant2016four, gies2018photon, king2018three},  in the sequel we focus on three-pulse collision schemes. Radiation of vacuum polarized by more than three pulses is also possible, but up to first order in $\alpha$ under consideration (see Eq.~\eqref{action}) \cite{klar2020detectable, gies2021quantum} is described by a combination of distributions for three-pulse schemes.

\subsection{Geometry of three-pulse collision schemes}
\begin{figure}[t]
\includegraphics[scale = 0.4]{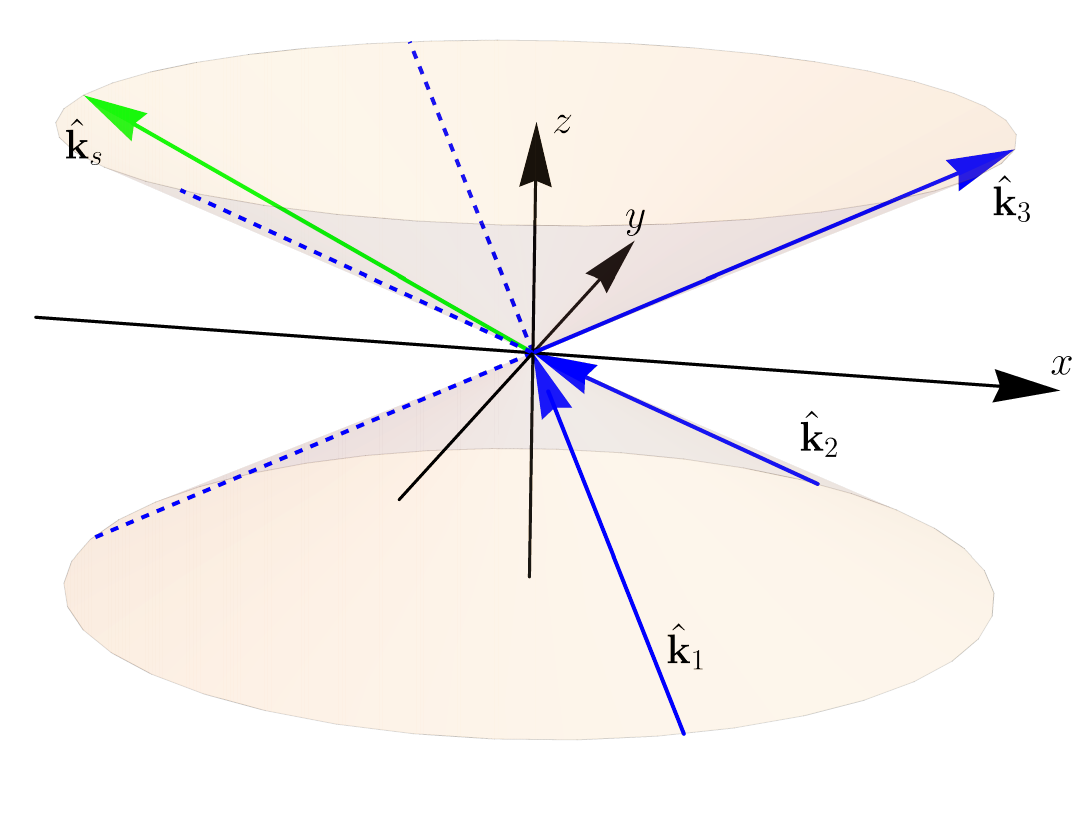}
\caption{\label{fig:geometry} A view of the collision scheme No.~7. The blue arrows indicate the propagation directions of the initial pulses, and the green arrow represents the signal photon.}
\end{figure}

First assume that all the three pulses are simultaneously focused precisely at the origin ($t_l = 0$, $\mathbf{r}_l = \mathbf{0}$) of the Cartesian coordinate system with the unit vectors~$\mathbf{\hat{e}}_x$,~$\mathbf{\hat{e}}_y$ and $\mathbf{\hat{e}}_z$. It is convenient to choose the direction of $\mathbf{\hat{e}}_z$ making the same acute angles with the directions of the pulses $\mathbf{\hat{k}}_1$, $\mathbf{\hat{k}}_2$ and $\mathbf{\hat{k}}_3$. Let us direct $\mathbf{\hat{e}}_x$ so that the vector~$\mathbf{\hat{k}}_3$ lies in the plane~$xz$, see Fig.~\ref{fig:geometry}. Then the directions of the pulses can be set with three parameters as follows:
\begin{subequations} \label{k123}
\begin{equation} \label{k12}
\mathbf{\hat{k}}_j = (\beta \cos \varphi_j, \, \beta \sin \varphi_j, \, \sqrt{1 - \beta^2}),  
\quad
j \in \{1, 2\},
\end{equation}
\begin{equation}
\mathbf{\hat{k}}_3 = (\beta, \, 0, \, \sqrt{1 - \beta^2}),
\end{equation}
\end{subequations}
where $\varphi_1$, $\varphi_2$ are the azimuth angles of the vectors $\mathbf{\hat{k}}_1$,~$\mathbf{\hat{k}}_2$,
and $0 \le \beta \le 1$ is the sine of the polar angle of each direction. In order to exclude from consideration the collision schemes differing by reflection in the $xz$ plane, we restrict the angle $\varphi_1$ to $0 \le \varphi_1 \le \pi$. Note that $\beta = 0$ corresponds to the case of codirectional pulses, and $\beta = 1$ to the case of all pulses propagating in the same plane.

The frequency $\omega_s$ and the direction $\mathbf{\hat{k}}_s$ of the signal photon are fixed by the conservation law
\begin{equation} \label{4-momentum}
k^\mu_1 + k^\mu_2 = k^\mu_3 + k^\mu_s, 
\end{equation}
where $k^\mu_l = (\omega_l, \mathbf{k}_l)$, $l \in \{1, 2, 3\}$; $k^\mu_s = (\omega_s, \mathbf{k}_s)$. The zero component of Eq.~\eqref{4-momentum} corresponds to energy conservation in photon-photon scattering and determines the frequency of the signal photon
\begin{equation} \label{frequencies}
\omega_s = \omega_1 + \omega_2 - \omega_3.
\end{equation}

The $z$-component of Eq.~\eqref{4-momentum} reads
\begin{equation} \label{z-momentum}
\omega_s \hat{k}_{s, \, z} = (\omega_1 + \omega_2 - \omega_3) \sqrt{1 - \beta^2}.
\end{equation}
By comparing it with Eq.~\eqref{frequencies},
we find $\hat{k}_{s, \, z} = \sqrt{1 - \beta^2}$, so that the direction of the signal photon can be sought in a form similar to Eq.~\eqref{k12}:
\begin{equation} \label{ks}
\mathbf{\hat{k}}_s = (\beta \cos \varphi_s, \,
\beta \sin \varphi_s, \, \sqrt{1 - \beta^2}).
\end{equation}

With account for Eq.~\eqref{ks}, the remaining components of Eq.~\eqref{4-momentum} take the form
\begin{eqnarray}
\notag
\omega_1 \cos \varphi_1 + \omega_2 \cos \varphi_2 &=& \omega_3 + \omega_s \cos \varphi_s, 
\\
\label{constraints}
\omega_1 \sin \varphi_1 + \omega_2 \sin \varphi_2 &=& \omega_s \sin \varphi_s.
\end{eqnarray}
Conditions \eqref{constraints} determine the azimuthal angle $\varphi_s$ of the signal photon, also imposing a restriction on the frequencies and directions of the initial pulses. Note that since Eqs.~\eqref{constraints} do not contain the parameter $\beta$, any combination of the parameters satisfying Eqs.~\eqref{constraints} defines a set of collision schemes differing by $\beta$, for $0\le\beta\le1$.

Since we are interested in optimizing the direction of propagation of the signal photon, consider the angles $\alpha_1$,~$\alpha_2$,~$\alpha_3$ ($0 \le \alpha_1, \alpha_2, \alpha_3 \le \pi$) that the pulses make with~$\mathbf{\hat{k}}_s$. Based on Eqs.~\eqref{k123}, \eqref{ks}, we find
\begin{eqnarray}
\notag
\sin \alpha_j &=& 2 \beta \left| \sin \frac{\varphi_j - \varphi_s}{2} \right|
\sqrt{1 - \beta^2 \sin^2 \frac{\varphi_j - \varphi_s}{2}}, 
\\
\label{sinaj}
\sin \alpha_3 &=& 2 \beta \left| \sin \frac{\varphi_s}{2} \right|
\sqrt{1 - \beta^2 \sin^2 \frac{\varphi_s}{2}},
\end{eqnarray}
where $j=1,2$.

A conceivable optimization target is making the deviation of the propagation direction of signal photons from all the incoming pulses as much as possible. To that end, it is enough to maximize the expression
\begin{equation} \label{max}
S = \min \, (\sin \alpha_1, \, \sin \alpha_2, \, \sin \alpha_3)
\end{equation}
under the constraints \eqref{constraints}. Finding a global maximum in the most general setting is a difficult task and requires a separate consideration. Assuming additionally that the incoming pulses are the harmonics $\omega_l=\nu_l\omega_0$ of a certain fundamental frequency~$\omega_0=2\pi c/\lambda_0$, the resulting (at least local) extrema of~$S$ are shown in Table~\ref{tab}. Note that scheme No.~8 does not maximize $S$ and is included for the sake of comparison as it has been extensively studied in the literature~\cite{lundstrom2006using,king2018three,gies2018photon,tennant2016four}.
Despite that all other listed schemes maximize angular separation $S$ of the signal, they might have certain  features unpleasant for experimental realization, such as head-on collision of some of the pulses (e.g. scheme No. 2), or that the frequency of signal photons coincides to some of the frequencies of the incoming pulses (as in scheme No. 5), or that higher than second harmonics are involved (e.g., scheme 15). Therefore in what follows we mostly have in mind scheme No. 7, which is slightly more favorable than the conventional scheme No. 8 in terms of both angular separation and signal photon yield.

\begin{table*}
\caption{\label{tab} Extrema of $S$ [see Eq.~\eqref{max}] for given pulse frequencies $\omega_l = \nu_l \omega_0$ with $\nu_s = \nu_1 + \nu_2 - \nu_3$.}
\begin{ruledtabular}
\begin{tabular}{ccccccccccc}
№ & $\nu_1$ & $\nu_2$ & $\nu_3$ 
& $\nu_s$ & $\varphi_1$ & $\varphi_2$ & $\varphi_s$ &  $\beta$ &
$S$ 
& $N_s\footnote{for parameters $P_l = 50 \, \text{PW}$, 
$\nu_l a_l = 0.0284$, $\nu_l \Delta_l = 0.159$, $\omega_0 = 2\pi c / \lambda_0$, 
$\lambda_0 = 910 \, \text{nm}$ and linear polarizations either $\theta_l = \pi / 4$ or $\theta_l = 3\pi / 4$ [see Sec.~\ref{sec4B} for details on the optimal choice of polarizations].}$
\\ \hline \\
1 & 
1 & 1 & 1 & 1 & 
$\pi/2$ & $3\pi/2$ & $\pi$ &
$\sqrt{2/3} \, (\approx 0.816)$ &
$\sqrt{8}/3 \, (\approx 0.943)$ 
& $3.03 \times 10^5$
\\
2 &
1 & 2 & 1 & 2 & 
$\pi$ & $\pi/3$ & $2\pi/3$ &
1 &
0.866
& $1.39 \times 10^6$
\\
3 &
& & & & 
2.30 & $3\pi/2$ & 3.82 &
0.954 &
0.784 
& $1.45 \times 10^6$
\\
4 &
& & & & 
1.86 & 4.49 & 3.66 &
0.955 & 
0.708 
& $2.00 \times 10^6$
\\
5 & 
1 & 2 & 2 & 1 & 
0.680 & $\pi$ & 1.91 &
1 &
$\sqrt{8}/3 \, (\approx 0.943)$ 
& $1.00 \times 10^6$
\\
6 &
& & & & 
0.876 & 5.25 & 4.38 &
0.933 &
0.727
& $2.38 \times 10^5$
\\
7 & 
2 & 2 & 1 & 3 & 
$2\pi/3$ & $4\pi/3$ & $\pi$ &
$2 / \! \sqrt{5} \, (\approx 0.895)$ & 
0.799
& $1.20 \times 10^6$
\\
8 & 
& & & & 
$2\pi/3$ & $4\pi/3$ & $\pi$ &
$\sqrt{2/3} \, (\approx 0.816)$ & 
$\sqrt{5} / 3 \, (\approx  0.745)$
& $5.67 \times 10^5$
\\
9 & 
&  &  &  & 
1.55 & 3.58 & 2.75 &
0.942 & 
0.705
& $1.82 \times 10^6$
\\
10 & 
&  &  &  & 
$\pi$ & 1.23 & 2.46 &
1 & 
0.629 
& $2.68 \times 10^6$
\\
11 & 
2 & 2 & 3 & 1 & 
$\pi/3$ & $5\pi/3$ & $\pi$ &
0.756 & 
0.990
& $1.00 \times 10^5$
\\
12 & 
&  &  &  & 
1.19 & 5.43 & 2.79 &
0.820 & 
0.952
& $1.79 \times 10^5$
\\
13 & 
&  &  &  & 
1.13 & 5.83 & 1.93 &
0.994 & 
0.712
& $2.99 \times 10^5$
\\
14 & 
2 & 3 & 1 & 4 & 
$\pi$ & 1.46 & 2.30 &
1 & 
$\sqrt{5} / 3 \, (\approx  0.745)$
& $1.84 \times 10^6$
\\
15 & 
&  &  &  & 
2.86 & 4.63 & 3.80 &
0.971 & 
0.722
& $1.83 \times 10^6$
\\
16 & 
&  &  &  & 
1.40 & 3.24 & 2.71 &
0.988 & 
0.501
& $3.61 \times 10^6$
\\
\end{tabular}
\end{ruledtabular}
\end{table*}

\subsection{Integrated signal}
Without loss of generality consider the harmonic $\mathbf{G}_{12}^3$. The corresponding number of signal photons is given by
\begin{equation} \label{Ns}
N_s = \int \frac{d \Omega_\mathbf{\hat{k}}}{4\pi}
\int\limits_{0}^{+\infty} \frac{\omega^3 d\omega}{\pi \hbar c^3}
 \left| \mathbf{C} \int d^4x A_1 A_2 A_3^*
e^{i(k - k_s)x} \right|^2,
\end{equation}
where
\begin{equation}
\mathbf{C} (\mathbf{\hat{k}}) = \mathbf{\hat{k}} \times \mathbf{p}_{12}^3 +
\mathbf{\hat{k}} \times \left( \mathbf{\hat{k}} \times \mathbf{m}_{12}^3 \right),
\end{equation}
and vector $\mathbf{m}_{12}^3$ is defined similarly to Eq.~\eqref{p123} as
\begin{equation} 
\mathbf{m}_{12}^3 \equiv 
e^{i (k_1 + k_2 - k_3)x}
\frac{\partial^3}{\partial A_1 \partial A_2 \partial A_3^*} 
\left( 4 \mathfrak{F} \mathbf{H} - 7 \mathfrak{G} \mathbf{E} \right).
\end{equation}
(see Appendix~\ref{app2} for the explicit expression).

The inner integral over spacetime in Eq.~\eqref{Ns} can be taken analytically using the infinite Rayleigh length approximation (IRLA) \cite{king2018three}. In doing so, we neglect the deviation of $\kappa_l$ from unity in Eq.~\eqref{alDeltal} (thus replacing~$\kappa_l \rightarrow 1$). This approximation is natural, since the region of three-pulse interaction is determined by the characteristic pulse focusing width~${\propto c / (\Delta \omega_0)}$, which is much smaller than the Rayleigh length ${\propto c / (\Delta^2 \omega_0)}$. In the IRLA approximation, the inner integral in Eq.~\eqref{Ns} becomes Gaussian, so that we find:
\begin{eqnarray} \label{Ns2}
N_s &\approx& \left( \frac{\alpha A_{0, \, 1} A_{0, \, 2} A_{0, \, 3}}{90 \pi E_c^2} \right)^2 \frac{(\nu_s c)^3}{\hbar \omega_0^4 \text{det} M} 
\nonumber
\\
& & \times \int d \Omega_\mathbf{\hat{k}}
\int\limits_0^{+\infty}  d\nu \, |\mathbf{C}|^2
\exp \left( - B^\text{T} M^{-1} B \right), 
\end{eqnarray}
where $\nu_s=\omega_s/\omega_0$ and the dimensionless column vector 
\begin{equation} \label{B}
B(\nu, \mathbf{\hat{k}}) = (\nu - \nu_s, -(\nu \mathbf{\hat{k}} - \nu_s \mathbf{\hat{k}}_s))^\text{T}
\end{equation}
depends on the state of the emitted signal photon, and
\begin{eqnarray}
M_{pq} = \left( \frac{c}{\omega_0} \right)^2 \frac{\partial^2}{\partial x^p \partial x^q} 
\sum_{l = 1}^3 \phi^2_l \left( t, M_l^{-1} \mathbf{r} \right),
\\
\phi_l^2 (t, \mathbf{r}) \equiv \phi_{\parallel, \, l}^2 + \phi_{\perp, \, l}^2
\end{eqnarray}
is dimensionless matrix of coefficients depending on the parameters $a_l$ and $\Delta_l$ (for an explicit expression see Appendix~\ref{app2}).

Using Eqs.~\eqref{Ns2},~\eqref{B}, we obtain an analytical expression for the angular distribution of the energy $W_s$ transferred to signal photons. Multiplying the integrand of Eq.~\eqref{Ns2} by the photon energy $\hbar \omega$, we arrive at
\begin{eqnarray} \label{dW}
\frac{d W_s}{d \Omega_{\hat{\mathbf{k}}}} &=& 
\left( \frac{\alpha A_{0, \, 1} A_{0, \, 2} A_{0, \, 3}}{90 \pi E_c^2} \right)^2 
\frac{|\mathbf{C}|^2 c^3}{\omega_0^3 \text{det} M} 
\nonumber
\\
& & \times 
\int\limits_0^{+\infty}  d\nu \nu^4
\exp \left( - B^\text{T} M^{-1} B \right).
\end{eqnarray}

The column vector $B$ \eqref{B} can be conveniently considered as a combination
\begin{equation}
B = \nu B_k - \nu_s B_s, 
\end{equation}
of the vectors
\begin{equation}
B_k \equiv (1, -\hat{\mathbf{k}})^\text{T}, 
\quad
B_s \equiv (1, -\hat{\mathbf{k}}_s)^\text{T},
\end{equation}
so that the argument of the Gauss exponential can be represented as
\begin{equation}
-B^\text{T} M^{-1} B  = 
- c_1 \nu^2 + c_2 \nu - c_3,
\end{equation}
where
\begin{eqnarray}
\notag
&& c_1 = B_k^\text{T} M^{-1} B_k, \\
\notag
&& c_2 = \nu_s \left( B_k^\text{T} M^{-1} B_s + B_s^\text{T} M^{-1} B_k \right), \\
\notag
&& c_3 = \nu_s^2 B_s^\text{T} M^{-1} B_s.
\end{eqnarray}

\begin{figure*}
\includegraphics[scale = 0.35]{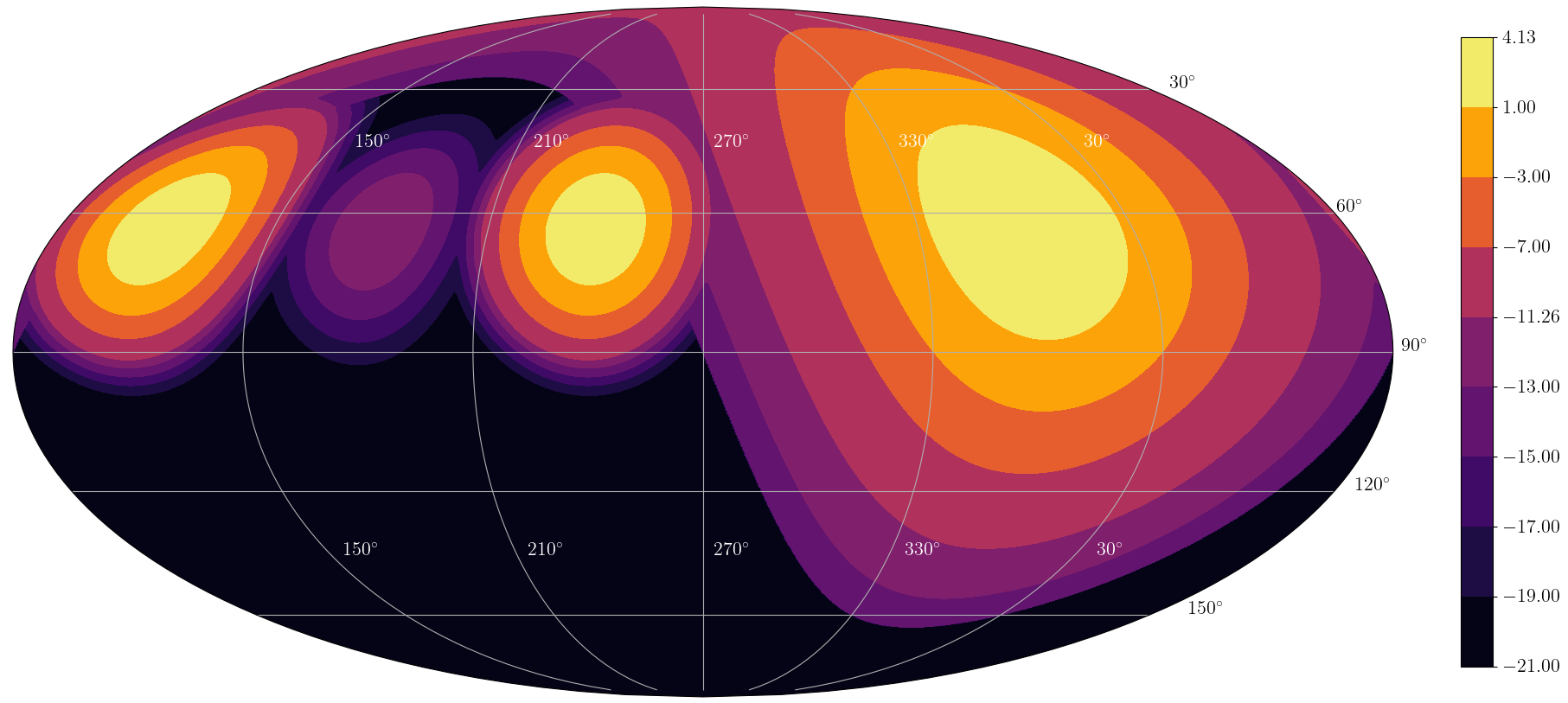}
\caption{\label{fig:angular_dist} Angular distribution Eq.~\eqref{dW_res} of the total energy of the signal photons in Joules (dimmer peak at about $\text{N} 64^o,\text{E}180^o$) compared to the energy distributions in initial pulses (brighter peaks, see Appendix~\ref{app1}) on a decimal logarithmic scale for scheme No.~7. All the three incoming pulses are polarized linearly ($\delta_l = 0$) with $\theta_l = \pi/4$, other parameters: $P_l = 50 \, \text{PW}$, 
$\nu_l a_l = 0.0284$, $\nu_l \Delta_l = 0.159$, $\omega_0 = 2\pi c / \lambda_0$, 
$\lambda_0 = 910 \, \text{nm}$.}
\end{figure*}

With the introduced notations, evaluation of the integral in Eq.~\eqref{dW} results in
\begin{widetext}
\begin{equation} \label{dW_res}
\frac{d W_s}{d \Omega_{\hat{\mathbf{k}}}} =
\left( \frac{\alpha  A_{0, \, 1} A_{0, \, 2} A_{0, \, 3}}{360 \pi E_c^3} \right)^2  
\frac{|\mathbf{C}|^2 (2 \mu (10 + \mu^2) + 
\sqrt{\pi} (12 + 12 \mu^2 + \mu^4) (1 + \text{erf} (\mu / 2)) e^{\mu^2 \! / 4}) e^{-c_3}}{2 c_1^{5/2} \text{det} M } \times
\frac{c^3 E_c^2}{\omega_0^3},
\end{equation}
\end{widetext}
where $\mu = c_2 / \! \sqrt{c_1}$. The corresponding signal energy distribution for scheme No.~7 from Table~\ref{tab}, combined with distributions for the incoming pulses, is shown in Fig. \ref{fig:angular_dist}, where it is clear that the signal, though much weaker, is nevertheless well separated from them.

Let us come back to the total number of the emitted signal photons. As already discussed, the integrand in Eq.~\eqref{Ns2} has a sharp peak at $\nu = \nu_s$ and $\mathbf{\hat{k}} = \mathbf{\hat{k}}_s$. Therefore, the integrals over the states of the emitted signal photon can be taken approximately using the saddle-point method.

Passing in Eq.~\eqref{Ns2} to the variables $(\nu, \theta, \varphi)$ of the spherical coordinate system, we have ${\mathbf{\hat{k}} = (\cos \varphi \sin \theta, \sin \varphi \sin \theta, \cos \theta)}$, and after evaluating the integrals to the lowest order obtain
\begin{eqnarray} \label{Ns3}
N_s &\approx& \left( \frac{\alpha A_{0, \, 1} A_{0, \, 2} A_{0, \, 3}}{90 E_c^2} \right)^2 
\nonumber
\\
& & \times \frac{ \beta (\nu_s c)^3 
|\mathbf{C}(\mathbf{\hat{k}}_s)|^2}{\sqrt{\pi} \hbar \omega_0^4 \text{det} M \sqrt{\text{det} \left( M_B^\text{T} M^{-1} M_B \right)}},   
\end{eqnarray}
where the matrix
\begin{equation}
M_B = \left. \left( \frac{\partial B}{\partial \nu}, \, \frac{\partial B}{\partial \theta}, \, 
\frac{\partial B}{\partial \varphi} \right)^\text{T} 
\right|_{\substack{\nu = \nu_s \\ \theta = \arcsin \beta \\ \phi = \phi_s}}
\end{equation}
is given explicitly in Appendix~\ref{app2}.

Finally, we rewrite Eq.~\eqref{Ns3} in terms of peak powers of the pulses using Eq.~\eqref{P},
thus arriving at
\begin{equation} \label{res}
N_s = N_0 \frac{\beta (\nu_1 \nu_2 \nu_3 \Delta_1 \Delta_2 \Delta_3)^2 \nu_s^3 C_s}{\text{det} M \sqrt{\text{det} \left( M_B^\text{T} M^{-1} M_B \right)}},
\end{equation}
where the following notations are introduced:
\begin{equation}
N_0 \equiv \frac{1}{\sqrt{\pi}} \left( \frac{32 \alpha}{45}\right)^2 \frac{\omega_0^2 P_1 P_2 P_3}{\hbar c^6 E_c^4}, \quad
C_s \equiv |\mathbf{C}(\mathbf{\hat{k}}_s)|^2.
\end{equation}
The quantity $N_0$ establishes the overall magnitude of the signal, whereas the remaining dimensionless factors in Eq.~\eqref{res} are scheme-specific corrections.

\subsection{Case of non-ideal collision of the pulses}
In deriving Eq.~\eqref{res} we assumed that the centers of all colliding laser pulses are simultaneously passing through the origin, and that their propagation directions satisfy the conservation laws Eq.~\eqref{4-momentum}. However, since such a perfect adjustment can hardly be achieved in practice, let us now refine Eq.~\eqref{res} for the case of shifted centers and directions of momenta of the pulses,
\begin{equation}
t_l \neq 0, \qquad \mathbf{r}_l \neq \mathbf{0}, \qquad k_l \rightarrow \tilde{k}_l = k_l + \delta k_l.
\end{equation}

Such changes obviously leave the inner four-dimensional integral in Eq.~\eqref{Ns} in the IRLA approximation Gaussian, so that it can still be taken analytically. An analogue of Eq.~\eqref{Ns2} takes the form
\begin{multline} \label{Ns4}
\tilde{N}_s \approx \left( \frac{\alpha A_{0, \, 1} A_{0, \, 2} A_{0, \, 3}}{90 \pi E_c^2} \right)^2 \frac{(\nu_s c)^3}{\hbar \omega_0^4 \text{det} \tilde{M}} 
\\
\times\!\int\!d \Omega_\mathbf{\hat{k}}
\int\limits_0^{+\infty}\!d\nu \, \left| \mathbf{C}
\exp \left( - \frac{\tilde{B}^\text{T} \tilde{M}^{-1} \tilde{B}}{2} - \frac{\gamma}{2} \right) \right|^2, 
\end{multline}
where
\begin{eqnarray}
\tilde{M}_{pq} = \left( \frac{c}{\omega_0} \right)^2 \frac{\partial^2}{\partial x^p \partial x^q} 
\sum_{l = 1}^3  \phi^2_l \left( \Delta t_l, \Delta \mathbf{r}_l \right),
\\
\gamma =
\sum_{l = 1}^3 \sum_{p, \, q = 0}^3 x_l^p x_l^q 
\frac{\partial^2 \phi^2_l \left( \Delta t_l, \Delta \mathbf{r}_l \right)}
{\partial x^p_l \partial x^q_l},
\\
\Delta t_l = t - t_l, \qquad \Delta \mathbf{r}_l = \left( M_l (\mathbf{\hat{\tilde{k}}}_l) \right)^{-1} \cdot (\mathbf{r} - \mathbf{r}_l),
\end{eqnarray}
and vector $\tilde{B}$ contains two corrections to $B$
\begin{equation} \label{tB}
\tilde{B}(\nu, \mathbf{\hat{k}}) = B (\nu, \mathbf{\hat{k}}) + \delta B^k - i \delta B^x, 
\end{equation}
the first one due to the change in the directions of the pulses
\begin{equation} 
\delta B^k_p = \left( \frac{c}{\omega_0} \right) \frac{\partial}{\partial x^p} \sum_{l = 1}^3 \delta k_l x,
\end{equation}
and the second one due to space and time shifts of their centers
\begin{equation}
\delta B^x_p = \left. \left( \frac{c}{\omega_0} \right) \frac{\partial}{\partial x^p} 
\sum_{l = 1}^3 \phi^2_l \left( \Delta t_l, \Delta \mathbf{r}_l \right) 
\right|_{\substack{t = 0 \\ \mathbf{r} = \mathbf{0}}}.
\end{equation}

Substituting Eq.~\eqref{tB} into Eq.~\eqref{Ns4}, we find:
\begin{widetext}
\begin{multline} \label{Nslast}
\tilde{N}_s \approx \left( \frac{\alpha A_{0, \, 1} A_{0, \, 2} A_{0, \, 3}}{90 \pi E_c^2} \right)^2 \frac{(\nu_s c)^3}{\hbar \omega_0^4 \text{det} \tilde{M}} 
\exp \left( (\delta B^x)^\text{T} \tilde{M}^{-1} \delta B^x - \gamma \right) 
\\
\times \int d \Omega_\mathbf{\hat{k}}
\int\limits_0^{+\infty}  d\nu \, |\mathbf{C}|^2
\exp \left( - \left( B + \delta B^k \right)^\text{T} 
\tilde{M}^{-1} \left( B + \delta B^k \right) \right).
\end{multline}
\end{widetext}

Note that the corrections due to a shift of the focal centers of the pulses are contained in Eq.~\eqref{Nslast} exclusively in the first exponential factor in front of the integral.

Further general analytical consideration of the expression \eqref{Nslast} is extremely bulky as the saddle point moves due to shifts in the directions of the pulses. Such a calculation is easier performed for particular cases using computer algebra.

\section{Results} \label{sec4}
Let us study the total number of signal photons in dependence on the parameters of a collision scheme. Due to the presence of a large number of parameters in  Eq.~\eqref{res}, it is worth restricting consideration to the most interesting special cases in Table~\ref{tab}. Namely, let us consider in more detail scheme No. 7, which is more favorable in terms of angular separation of signal photons than the conventional scheme No. 8. These two schemes are specified by the following parameter values:
\begin{equation}
\nu_1 = \nu_2 = 2 \nu_3 = 2, \quad
2\varphi_1 = \varphi_2 = \frac{4\pi}{3}, \quad \varphi_s = \pi.
\end{equation}
In this case, specified also to coinciding durations and focal widths, and linear polarizations of the pulses
\begin{equation} \label{equal_pulses}
\nu_l a_l = a, \qquad \nu_l \Delta_l = \Delta, \qquad \delta_l = 0,
\end{equation}
Eq.~\eqref{res} takes the form
\begin{widetext}
\begin{equation} \label{Ns7}
N_s = N_0
\frac{\Delta^5 C_s}{6 \sqrt{3} \beta^2 a (\beta^2 a^2 + (2 - \beta^2) \Delta^2) \sqrt{3\beta^2 a^2 + (4 - 3\beta^2) \Delta^2}},
\end{equation}
\begin{eqnarray} \label{Cs}
\left. C_s \right|_{\delta_l = 0} &=& \frac{9 \beta^8}{2048} 
\left( 1251 + 145 \cos(2 (\theta_1 - \theta_2)) - 
816 \cos(2 (\theta_1 + \theta_2)) \right. \notag
\\
&& + \left. 515( \cos(2 (\theta_1 - \theta_3)) + \cos(2 (\theta_2 - \theta_3))) 
- 229( \cos(2 (\theta_1 + \theta_3)) + \cos(2 (\theta_2 + \theta_3)) )
\right).
\end{eqnarray}
\end{widetext}

\begin{figure*}[t!]
\begin{minipage}[ht]{0.49\linewidth}
\center{\includegraphics[scale = 0.5]{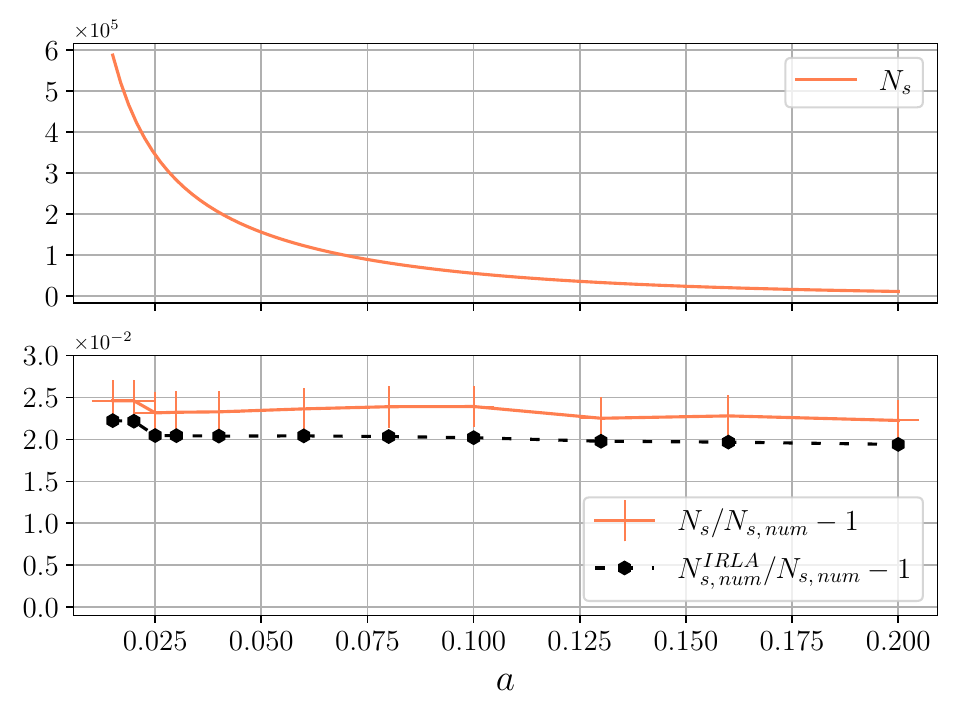}}
\end{minipage}
\hfill
\begin{minipage}[ht]{0.49\linewidth}
\center{\includegraphics[scale = 0.5]{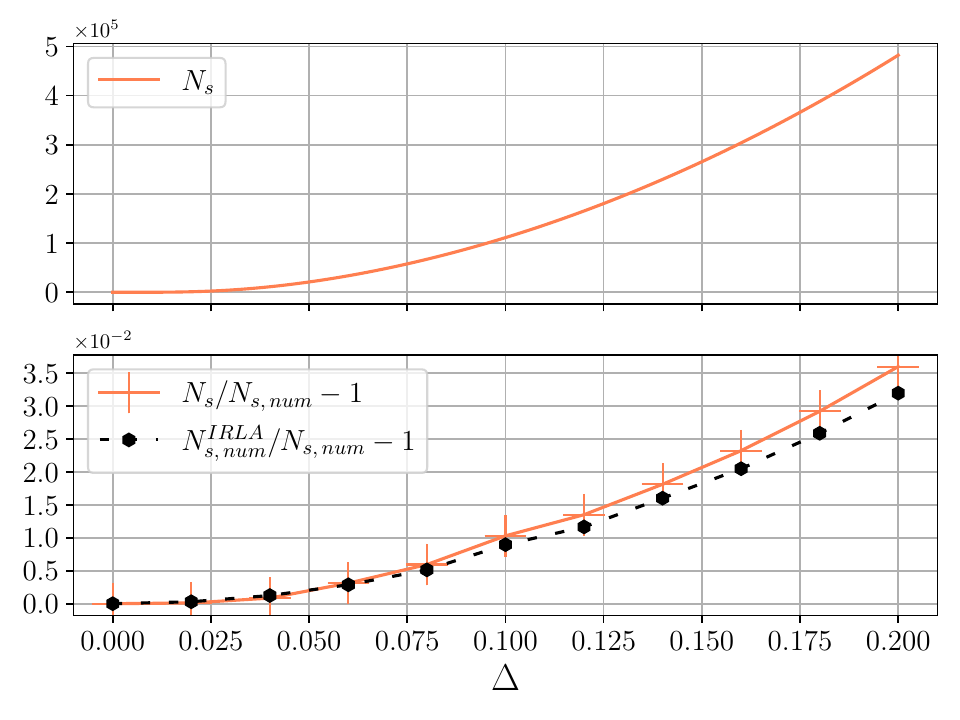}}
\end{minipage}
\caption{\label{fig:Ns(D)} The total number of signal photons and its relative error for collision scheme No. 7 calculated using analytical formula~\eqref{res}, as well as using numerically in IRLA approximation, against numerical evaluation without approximations versus the pulses duration (left) and focusing width (right). Parameters: $\theta_l = \pi/4$, $\delta_l = 0$, $\Delta = 0.159$ (left), $a = 0.0284$ (right).}
\end{figure*}

In what follows, for definiteness we focus on the particular parameter values $\lambda_0 = 910 \, \text{nm}$ and ${2P_1 = 2P_2 = P_3 = 50 \, \text{PW}}$, such that $N_0 = 7 \times 10^5$. This roughly corresponds to the parameters opted for the XCELS facility \cite{khazanov2023exawatt} with account for typical conversion efficiency of second harmonics generation $\sim 50\%$ \cite{marcinkevivcius2004frequency, dansette2020peculiarities,andral2022second}. Since the dependence on these parameters in Eq.~\eqref{res} is trivial, rescaling of our results for other facilities is straightforward.

As for other parameters, the adopted values $a = 0.0284$ and $\Delta = 0.159$ correspond to pulse duration $\tau = 17 \, \text{fs}$ (so that FWHM of the pulse energy distribution is equal to~$20 \, \text{fs}$) and focal width ${w = \lambda_0}$ (see Appendix~\ref{app1}).

\subsection{Accuracy of approximations in dependence on pulse width and duration}

To obtain Eq.~\eqref{res} we applied the IRLA approximation along with the saddle-point method. Now let us explore the accuracy of the result by comparing Eq.~\eqref{res} with direct numerical evaluation of the integral. The results of such comparison given in Fig.~\ref{fig:Ns(D)} show that the saddle-point method is a weaker approximation than IRLA, for which the error scales as $\Delta^2$. Therefore, to refine the result, one first needs to take into account the longitudinal component of the field in the pulse model, linear in~$\Delta$. Without taking this into account the total error does not exceed 4\% in the considered range of parameters. In view of the above, all further consideration will be based on the obtained analytical formula.

\subsection{Dependence on pulses polarization}\label{sec4B}

\begin{figure}[b]
\includegraphics[scale = 0.5]{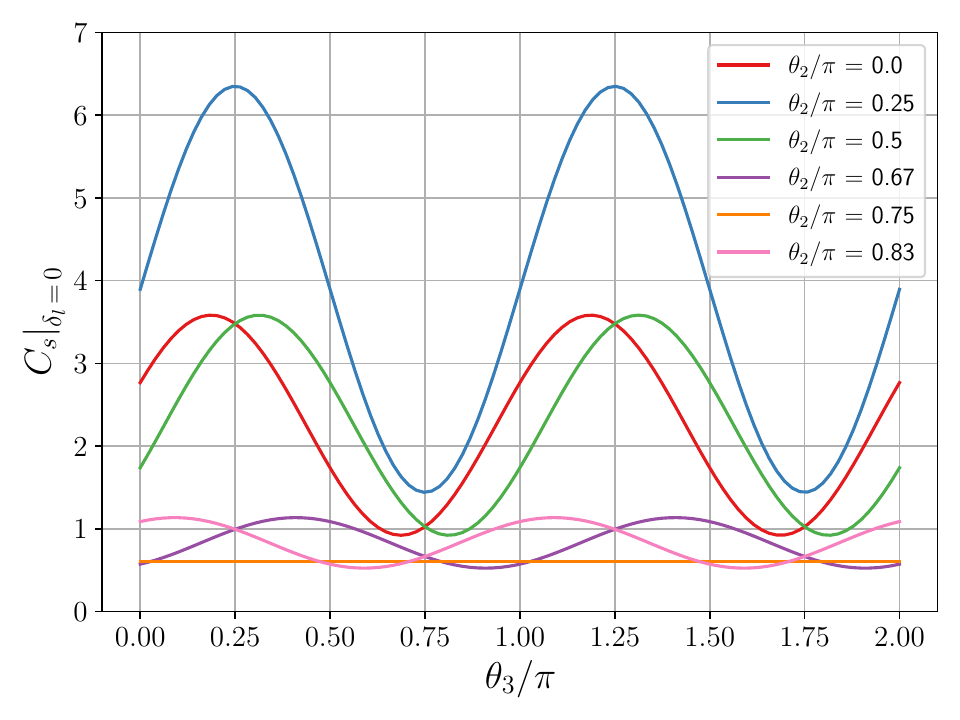}
\caption{\label{fig:Ns(th3)} Dependences of the polarization factor $C_s$ for collision scheme No. 7 with $\beta = 2 / \! \sqrt{5}$ in the case of linear polarizations ($\delta_l = 0$) on the polarization angle $\theta_3$ of the third pulse for different values of the polarization angle $\theta_2$ of the second pulse for the fixed value $\theta_1 = \pi / 4$ of the polarization angle of the first pulse.}
\end{figure}

Let us investigate the dependence of the factor $C_s$ on the pulses polarization in scheme No. 7. Consider first the case of linear polarization \eqref{Cs}. The  dependence on the polarization angles $\theta_l$ is shown in Fig.~\ref{fig:Ns(th3)}. In particular,~\eqref{Cs} attains maximum
\begin{equation} \label{C_lin_max}
\left. \max C_s \right|_{\delta_l = 0} = \left( \frac{63}{16} \right)^2 \beta^8 
\approx 15.5 \, \beta^8 
\end{equation}
at $\theta_l = \pi / 4$ or $\theta_l = 3\pi / 4$. It turns out that these polarization angles actually maximize the signal for all the schemes in Table~\ref{tab}.

\begin{figure}[t]
\includegraphics[scale = 0.5]{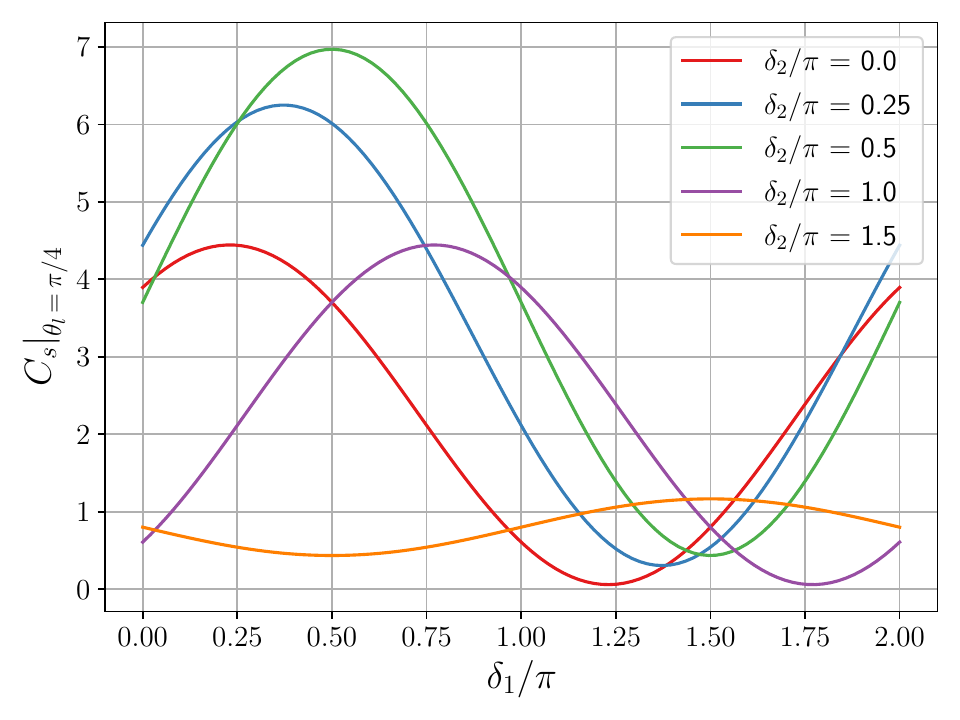}
\caption{\label{fig:Ns(d1)} Dependences of the polarization factor $C_s$ for collision scheme No. 7  with $\beta = 2 / \! \sqrt{5}$ on the ellipticity $\delta_1$ of the first pulse ($\theta_1 = \pi / 4$) for different values of the ellipticity $\delta_2$ of the second pulse ($\theta_2 = \pi / 4$) for a fixed polarization $\delta_3 = 3\pi / 2$, $\theta_3 = \pi/4$ of the third pulse.}
\end{figure}

We note that in general the parameter $\beta$ factors out in the expression for $C_s$, which, in this case, has a maximum
\begin{equation}
\max C_s = \left( \frac{33}{8} \right)^2 \beta^8 
\approx 17.0 \, \beta^8 
\end{equation}
at $(\theta_l, \, \delta_l) \in \{ (\pi/4, \pi/2), (3\pi/4, 3\pi/2) \}$ or
$(\theta_l, \, \delta_l) \in \{ (\pi/4, 3\pi/2), (3\pi/4, \pi/2) \}$, i.e. the maximum is attained for circular identically oriented polarizations of the pulses. This maximum, however, only slightly exceeds Eq.~\eqref{C_lin_max}. Dependence of the factor~$C_s$ on the ellipticities~$\delta_l$ is shown in Fig.~\ref{fig:Ns(d1)}. 

The same analysis applies to the dependence on polarization for scheme No. 8.

\subsection{Dependence on scheme planarity}

\begin{figure}[t]
\includegraphics[scale = 0.5]{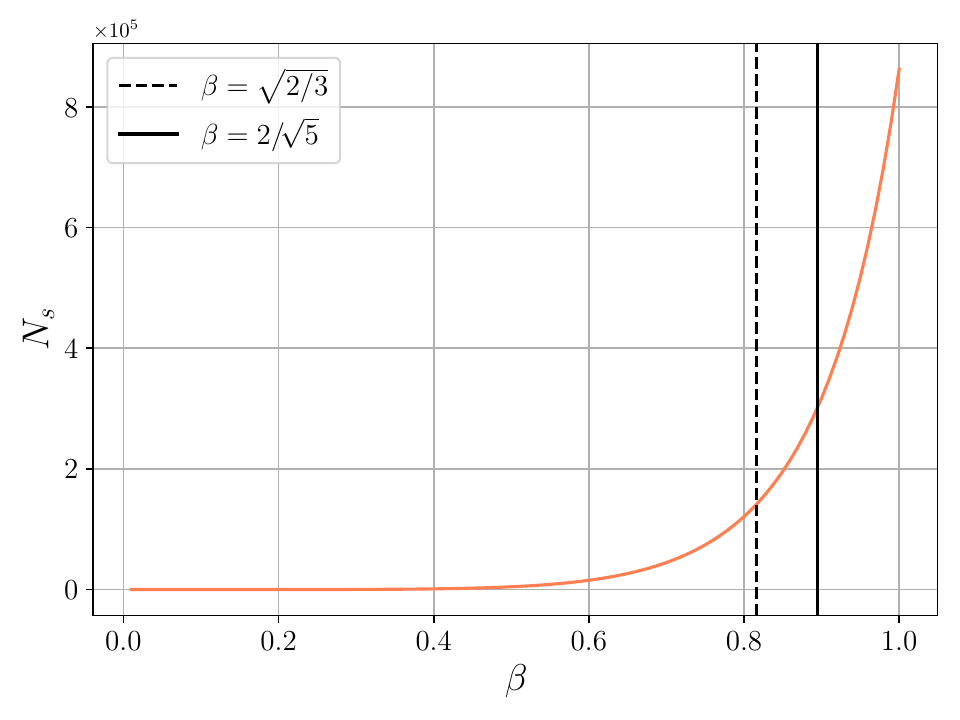}
\caption{\label{fig:Ns(beta)} Dependence of the number of signal photons \eqref{Nslast} on the parameter $\beta$ of the collision geometry. Black solid and dashed lines correspond to schemes No. 7 and 8, respectively. Parameters: $a = 0.0284$, $\Delta = 0.159$, $\theta_l = \pi / 4$, $\delta_l = 0$.}
\end{figure}

Let us study the dependence of the number of signal photons on the parameter $\beta$ of the collision geometry. This parameter is the sine of the polar angle of the directions of both the initial pulses and signal photons, thereby it sets the planarity degree  of the collision geometry. In particular, we note that  schemes Nos. 7 and 8 differ only by the value of $\beta$. The dependence of the number of signal photons on $\beta$ is shown in Fig.~\ref{fig:Ns(beta)}. The number of signal photons increases with $\beta$, as it also follows from Eqs.~\eqref{Nslast}, \eqref{Cs}. At the same time, the number of signal photons in scheme No. 7 is 2.15 times greater than in scheme No. 8. Further increase in $\beta$, though leads to an increase in the number of signal photons, worsens  spatial separation between the signal and the initial pulses. For example, at $\beta = 1$ the direction of the signal photons is opposite to the third pulse.

\subsection{Dependence on widths difference of the pulses}

\begin{figure*}
\begin{minipage}[ht]{0.49\linewidth}
\center{\includegraphics[scale = 0.55]{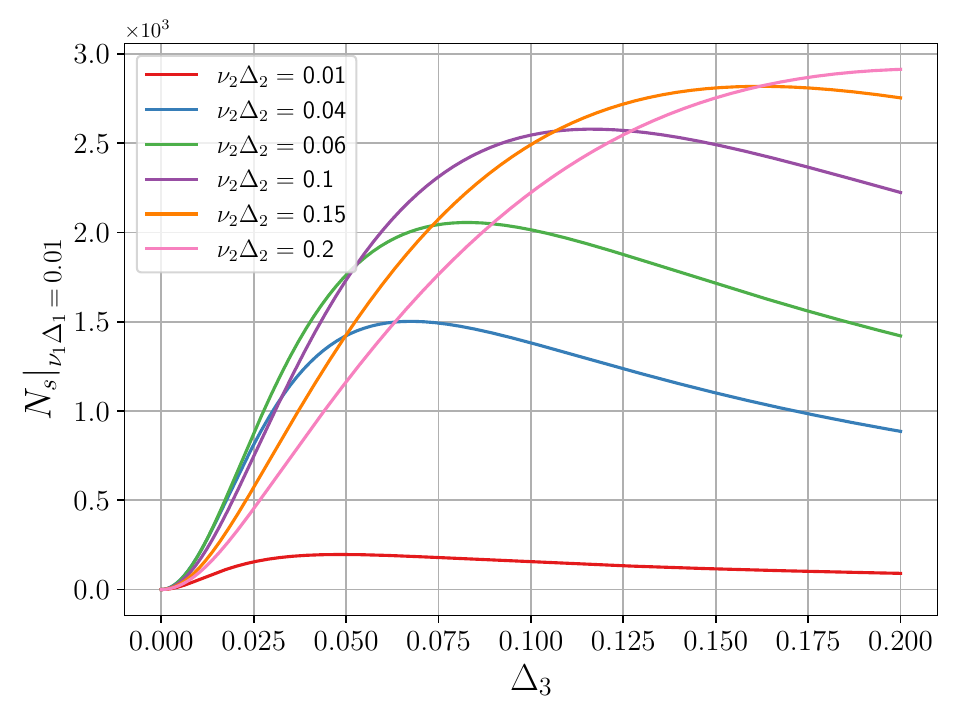}}
\end{minipage}
\hfill
\begin{minipage}[ht]{0.49\linewidth}
\center{\includegraphics[scale = 0.55]{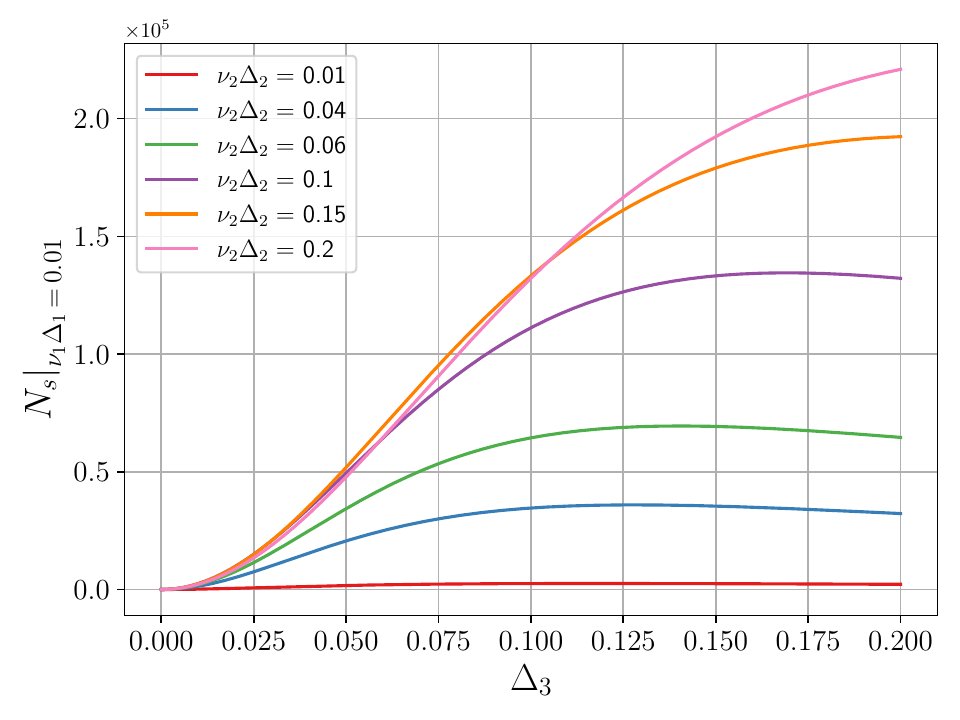}}
\end{minipage}
\caption{\label{fig:Ns(D3)} The dependence of the number of signal photons in scheme No. 7 on the  focusing parameter~$\Delta_3$ of the third pulse for different values of the spatial focusing parameter $\Delta_2$ of the second pulse and fixed value $\Delta_1 = 0.005$ (left) and $\Delta_1 = 0.05$ (right) of the focusing parameter of the first pulse. Other parameters: $a = 0.0284$, $\theta_l = \pi / 4$, $\delta_l = 0$.}
\end{figure*}

The dependence of the signal on the focusing width of the third pulse for different values of the focusing width of the second pulse and fixed focusing width of the first pulse are shown in Fig. \ref{fig:Ns(D3)}.

It follows that the magnitude of the signal is mainly determined by the least focused pulse: the smaller its focusing width, the stronger the effect. As long as the pulse focusing widths remain commensurable, strengthening the focusing of any of the pulses leads to an increase in the number of signal photons, otherwise further focusing of the most focused pulse only weakens the effect. Thus, the overall effect is optimized by strongest possible focusing of all the pulses, keeping their focusing commensurable, but not necessarily precisely the same.

\subsection{Comparison with previous studies}
In comparing formula \eqref{res} with the results of previous studies there is a number of difficulties related to different choices of a model for the laser pulses and the discrepancies on the definitions of its parameters. For example, Ref.~\cite{lundstrom2006using} considered scheme No.~8 but for simplicity with flat top profiles of the overlapping pulses. Therefore, a literal comparison of the number of signal photons is hardly possible. Nevertheless, for parameters $\lambda_0 = 800\,\text{nm}$, $5P_1 = 5P_2 = P_3 = 0.5\,\text{PW}$, $c \tau_l = 10\,\text{$\mu$m}$, optimal linear polarizations \eqref{C_lin_max} and assuming the focus width is half of the spatial interaction region $w_l = \lambda_0$, we get 0.067 signal photons which agrees well with 0.07 photons in Ref.~\cite{lundstrom2006using}. 
More important is the agreement in  dependence on the parameters of the pulses. In particular, the dependence on the duration $\tau$ and focusing width $w$, which from our Eq.~\eqref{Ns7} in case $a \ll \Delta$ under consideration reads
\begin{equation}
N_s \sim \frac{\Delta^2}{\lambda_0^2 a} \sim \frac{\tau}{\lambda_0 w^2},
\end{equation}
agrees with Eq.~8 in Ref.~\cite{lundstrom2006using} if one assumes that $\tau\sim L$ and $w\sim \lambda_0$. As for the dependence \eqref{Cs} on the polarization angle, our Fig.~\ref{fig:Ns(th3)} up to a horizontal shift and vertical stretch coincides to the upper panel in Fig.~2 of Ref.~\cite{lundstrom2006using}.

Later works have already used the Gaussian beam model which up to the notation is the same as \eqref{beaml}-\eqref{alDeltal}. However, Ref.~\cite{king2018three} does not indicate explicitly the relationship between the amplitude and power of the pulses, as well as the assumed polarization angles of the incoming pulses, so that it is impossible to accurately restore the setup. For parameters $\lambda_0 = 910\,\text{nm}$, $P_l = 25\,\text{PW}$, $\tau_l = 30\,\text{fs}$, $w_l = 5\,\text{$\mu$m}$ and optimal linear polarizations~\eqref{C_lin_max} we obtain 3200 signal photons, as opposed to~760 in Ref.~\cite{king2018three}. On the other hand, we also observe that IRLA overestimates the number of signal photons (see Fig. \ref{fig:Ns(D)}).

In Ref.~\cite{gies2018all}, the authors also use the Gaussian beam model but carefully indicate all the required pulses parameters. Setting $\lambda_0 = 800\,\text{nm}$, $2A_{0,\,1} = 2A_{0,\,2} = A_{0,\,3} = 8.26 \times 10^{-3} E_c$ (these amplitudes correspond to the energies $6.25/6.25/25 \, J$ chosen in Ref.~\cite{gies2018all} with account for the connection given there and the application of different system of units for the fields, which effectively leads to an additional factor $\sqrt{4\pi}$), $\tau_l = 12.5\,\text{fs}$ (after bisection due to different notations), $w_l = \lambda_0$ and optimal linear polarizations \eqref{C_lin_max}, we arrive at 1.23 photons, which is about twice less than 2.42 photons as indicated in Ref.~\cite{gies2018all}.

\subsection{Accounting for non-ideal setting of the experiment}
\begin{figure}[b]
\includegraphics[scale = 0.5]{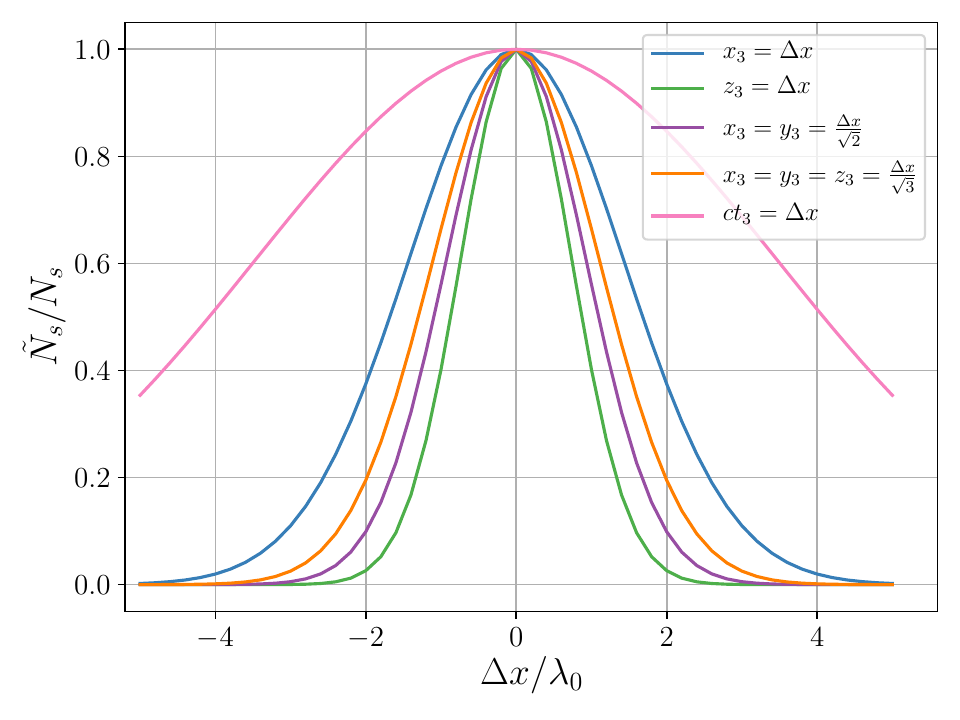}
\caption{\label{fig:Ns(x3)} 
The ratio $\Tilde{N}_s/N_s=\exp(-\gamma)$, see Eq.~\eqref{Nslast}, of the numbers of signal photons for the cases of a displaced focal center of the third pulse and exact convergence. 
\\
Parameters: ${a = 0.0284}$, ${\Delta = 0.159}$.}
\end{figure}
\begin{figure}[t]
\includegraphics[scale = 0.5]{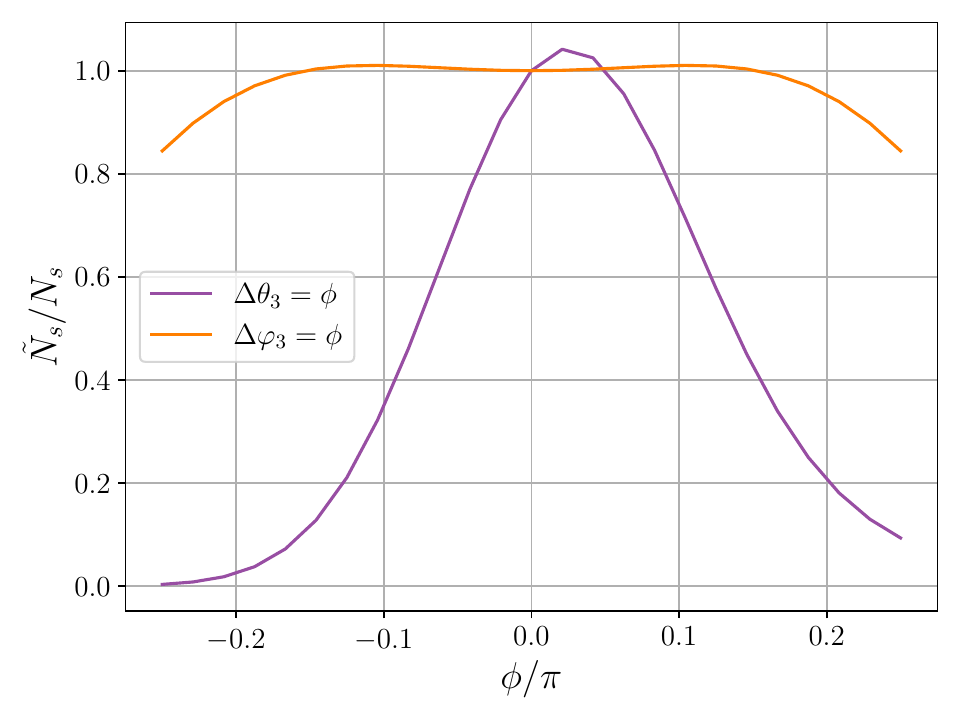}
\caption{\label{fig:Ns(angles3)} The ratio $\Tilde{N}_s/N_s$ of the numbers of signal photons for the cases of shifted (see Eq.~\eqref{Nslast}) and initial $\mathbf{\hat{k}}_3$ propagation directions  of the third pulse in scheme No.~7. The two curves correspond to shifts in either polar or azimuth angle with respect to the initial direction. 
\\
Parameters: $a = 0.0284$, $\Delta = 0.159$, $\theta_l = \pi / 4$, $\delta_l = 0$.}
\end{figure}

Non-ideal settings include non-perfect spatial and temporal matching of the incoming pulses and imperfection of the vacuum in the overlapping region. The former can be estimated using Eq.~\eqref{Nslast}.

The relative decrease in the signal photons yield due to a mismatch of the focal centers of the incoming pulses  is shown in Fig.~\ref{fig:Ns(x3)}. For definiteness we perturb scheme No.~7 assuming deviation of pulse $L_3$ either in focal center or in propagation direction. 

Since for the assumed parameters the length of the pulses is much greater than their focal width ($a \ll \Delta$), as expected, the effect is least sensitive to the mismatch in time overlap, but most sensitive to the accuracy of spatial overlap.

The reduction of the number of signal photons due to a shift in the direction of propagation of the third pulse is shown in Fig.~\eqref{fig:Ns(angles3)}. Here the effect is much more sensitive to the changes of the polar angle of $\mathbf{\hat{k}}_3$ than to the azimuthal one. This can be understood by that a change in the polar angle necessarily leads to the violation of the equality \eqref{z-momentum}, thereby violating the 4-momentum conservation law, whereas for a change of the azimuthal angle, the conservation laws~\eqref{constraints} remain satisfied with the direction of emission of signal photons shifted correspondingly.

The closest in magnitude competitor  to photon-photon scattering is Compton scattering of laser pulses by residual electrons. To reduce the corresponding noise, it is enough to ensure a sufficiently low pressure in the vacuum chamber. A rough estimate of the required vacuum purity can be obtained by the requirement that the residual electron current $j_e = en v$ (where $n$ and $v$ are the residual electron density and velocity) is smaller than the vacuum current \eqref{sources}. This way the pressure should obey
\begin{equation}
p \le 10^{-6} \, \text{mbar}
\end{equation}
for sufficient suppression of the effect.

\section{Conclusion} \label{sec5}

The paper examines experimental setup for detecting real photons emitted in a three-pulse collision due to photon-photon scattering. Our approach is based on an approximate analytical formula for the signal photons yield. This formula is derived assuming the Gaussian beam model by isolating the weakly oscillating contribution corresponding to the fulfillment of energy-momentum conservation law, applying the IRLA and using the saddle point method. The resulting formula includes the dependence on the geometry of the collision and on the set of parameters for each pulse, such as power, arbitrary polarization, and focusing duration and width parameters. Though it looks extremely complicated in general setting, it proved to be immensely useful for the analysis and optimization. We estimate its relative error as not exceeding $4\%$ in a wide range of parameters. In principle, our approach admits further generalizations, e.g. on higher number of interacting pulses and by taking into account next to leading orders in the mentioned approximations. However, we have shown that the main inaccuracy should come from the non-paraxial corrections to the used pulse model.

One of the goals of the paper was optimization of the collision scheme.
We demonstrate that the closer the arrangement of the incoming pulses to a single plane, the stronger is the signal for fixed other parameters. At the same time, there exists an optimal planarity in terms of angular separation of the signal. We propose a number of collision schemes that maximize angular separation between the signal and the incoming pulses. One of them is close to the one that was extensively studied in the literature but provides better angular separation and stronger signal. We analyzed this new scheme in detail. In particular, we have confirmed strong dependence on polarization of the pulses and identified their circular polarization as the optimal one. We have showed that the signal is stronger for stronger focused pulses with commensurable widths.  

We have also studied the effect of non-ideal settings. For realistic settings, it turns out that the precision of spatial overlap of the incoming pulses is more crucial than of their temporal overlap. Also, it is important to direct the pulses so to maintain the planarity of the scheme. The purity of the required vacuum roughly corresponds to the parameters of the vacuum chambers installed at the state-of-the-art laser facilities.

For definiteness, the particular numerical estimates were mostly made for the parameters of the XCELS project \cite{khazanov2023exawatt}. However, general formulas should apply to planning any experiment of this kind based on three-wave collision.

\begin{acknowledgments} 
The authors are greatful to I.Yu. Kostyukov for valuable discussions. This work was supported by the scientific program of the National Center for Physics and Mathematics, the MEPhI Program Priority 2030 and the Russian Foundation for Basic Research (project No.~20-52-12046, accounting for non-ideal setting of the experiment).
	
\end{acknowledgments}

\appendix
\section{Relationship between the parameters of the Gaussian pulse model}
\label{app1}
Let us express the parameters $a$, $\Delta$, and $A_0$ of our focused laser pulse model \eqref{beaml}-\eqref{alDeltal} in terms of conventional ones.

Parameter $a$ can be expressed in terms of pulse duration $\tau$ by comparing the temporal envelope shape:
\begin{equation} \label{temporal envelop}
\exp \left( -a^2 \omega^2 \, t^2 \right)
\equiv \exp \left( -\frac{t^2}{\tau^2} \right).
\end{equation} 
From \eqref{temporal envelop} we get
\begin{equation} \label{a}
a = \frac{1}{\omega \tau} = 
\frac{\lambda}{2 \pi c \tau}.
\end{equation}

It is natural to express parameter $\Delta$ similarly to \eqref{temporal envelop}, \eqref{a} in terms of focusing width $w$, which leads to
\begin{equation} 
\Delta = \frac{c}{\omega w} = 
\frac{\lambda}{2 \pi w}.
\end{equation}

Parameter $A_0$, which is proportional to the beam amplitude, can be expressed in terms of peak power $P$ or total energy $W$ of the pulse. Gaussian beam power
\begin{eqnarray}
\frac{c}{8 \pi} \int \! dxdy \, \text{Re} \left( \tilde{\mathbf{E}}^L_s \times \left( \tilde{\mathbf{H}}^{L}_s \right)^* \right) \notag
\\
= \frac{c^3 
A_0^2 \, e^{-2a^2\omega^2(t-z)^2}}{16 \Delta^2 \omega^2}
\end{eqnarray}
attains maximum $P$ at its center $z = t$, hence
\begin{equation} \label{P}
A_0 = \frac{4 \Delta \omega}{c} \sqrt{\frac{P}{c}} = 
\frac{4}{w} \sqrt{\frac{P}{c}} \, .
\end{equation}

On the other hand, the total beam energy
\begin{eqnarray}
\notag
W &=& \frac{c}{8 \pi} \int \! dtdxdy \, \text{Re} \left( \tilde{\mathbf{E}}^L_s \times \left( \tilde{\mathbf{H}}^{L}_s \right)^* \right)
\\
&& = \frac{\sqrt{\pi} c^3 A_0^2}{16 \sqrt{2} \, a \Delta^2 \omega^3}, 
\end{eqnarray}
therefore
\begin{equation}
A_0 = 4 \sqrt{\sqrt{\frac{2}{\pi}} \, \frac{a \Delta^2 \omega^3 W}{c^3}} =
4 \sqrt{\sqrt{\frac{2}{\pi}} \, \frac{W}{c \tau w^2}} \, .
\end{equation}

Note also that the angular distribution of the total energy of the Gaussian beam 
\begin{equation}
\frac{d W}{d \Omega} = 
\left. \left( \frac{c R^2}{8 \pi} \int \! dt \, \text{Re}
\left( \tilde{\mathbf{E}}^L_s \times 
\left( \tilde{\mathbf{H}}^{L}_s \right)^* \right) \right)
\right|_{\substack{z\approx \theta R, \\ R \rightarrow \infty}}
\end{equation}
has the form
\begin{equation}
\frac{d W}{d \Omega} = 
\frac{P e^{-\theta^2 \! / \Delta^2}}{2 \sqrt{2 \pi} \, a \Delta^2 \omega} = 
\frac{W e^{-\theta^2 \! / \Delta^2}}{2 \pi \Delta^2},
\end{equation}
where $\theta$ is the polar angle of the coordinate system.

\begin{widetext}
\section{Explicit form of the vectors $\mathbf{p}_{12}^3$, $\mathbf{m}_{12}^3$ and matrices $M$ and $M_B$} \label{app2}
Components of the complex vector $\mathbf{p}_{12}^3$ in terms of the polarization vectors $\mbox{\boldmath $\epsilon$}_l = \left( \epsilon_{l, \, x}, \, \epsilon_{l, \, y}, \, 0 \right)$ (in reference frames of the pulses) and geometry parameters of the scheme are given by: 
\begin{eqnarray}
\notag
p_{12, \, x}^3 &=&
\beta^2 (2  \epsilon_{3, \, y}^* (4 ( \epsilon_{1, \, x}  \epsilon_{2, \, x} -  \epsilon_{1, \, y}  \epsilon_{2, \, y}) \sqrt{1 - \beta^2} + 7  \epsilon_{1, \, y}  \epsilon_{2, \, x} \sin{\varphi_1} - 4  \epsilon_{1, \, x}  \epsilon_{2, \, y} \sin{\varphi_1} 
\\
\notag
&-& \cos{\varphi_2} ((7  \epsilon_{1, \, x}  \epsilon_{2, \, x} + 4  \epsilon_{1, \, y}  \epsilon_{2, \, y}) \sqrt{1 - \beta^2} + (7  \epsilon_{1, \, y}  \epsilon_{2, \, x} - 4  \epsilon_{1, \, x}  \epsilon_{2, \, y}) \sin{\varphi_1}) - 4  \epsilon_{1, \, y}  \epsilon_{2, \, x} \sin{\varphi_2} + 7  \epsilon_{1, \, x}  \epsilon_{2, \, y} \sin{\varphi_2} 
\\
\notag
&-& 4  \epsilon_{1, \, x}  \epsilon_{2, \, x} \sqrt{1 - \beta^2} \sin{\varphi_1} \sin{\varphi_2} + 
4  \epsilon_{1, \, y}  \epsilon_{2, \, y} \sqrt{1 - \beta^2} \sin{\varphi_1} \sin{\varphi_2} + \cos{\varphi_1} (-((7  \epsilon_{1, \, x}  \epsilon_{2, \, x} + 4  \epsilon_{1, \, y}  \epsilon_{2, \, y}) \sqrt{1 - \beta^2}) 
\\
\notag
&+& 2 (5  \epsilon_{1, \, x}  \epsilon_{2, \, x} + 6  \epsilon_{1, \, y}  \epsilon_{2, \, y}) \sqrt{1 - \beta^2} \cos{\varphi_2} + 
(4  \epsilon_{1, \, y}  \epsilon_{2, \, x} - 7  \epsilon_{1, \, x}  \epsilon_{2, \, y}) \sin{\varphi_2})) 
\\
\notag
&+&  \epsilon_{3, \, x}^* (-14 ( \epsilon_{1, \, y}  \epsilon_{2, \, x} +  \epsilon_{1, \, x}  \epsilon_{2, \, y}) \sqrt{1 - \beta^2} 
 + 2 (4  \epsilon_{1, \, y}  \epsilon_{2, \, x} - 7  \epsilon_{1, \, x}  \epsilon_{2, \, y}) \sqrt{1 - \beta^2} \cos{\varphi_1}
\\
\notag
&+& 17 ( \epsilon_{1, \, y}  \epsilon_{2, \, x} +  \epsilon_{1, \, x}  \epsilon_{2, \, y}) \sqrt{1 - \beta^2} \cos(\varphi_1 - \varphi_2) - 14  \epsilon_{1, \, y}  \epsilon_{2, \, x} \sqrt{1 - \beta^2} \cos{\varphi_2} 
\\
\notag
&+& 8  \epsilon_{1, \, x}  \epsilon_{2, \, y} \sqrt{1 - \beta^2} \cos{\varphi_2} + 3  \epsilon_{1, \, y}  \epsilon_{2, \, x} \sqrt{1 - \beta^2} \cos{(\varphi_1 + \varphi_2)} + 3  \epsilon_{1, \, x}  \epsilon_{2, \, y} \sqrt{1 - \beta^2} \cos{(\varphi_1 + \varphi_2)} 
\\
&+& 2 (4  \epsilon_{1, \, x}  \epsilon_{2, \, x} + 7  \epsilon_{1, \, y}  \epsilon_{2, \, y}) (\sin{\varphi_1} + \sin{\varphi_2} - \sin{(\varphi_1 + \varphi_2)})))/16,
\\
\notag
\\
\notag
p_{12, \, y}^3 &=& \beta^2 ( \epsilon_{3, \, x}^* (4  \epsilon_{1, \, y}  \epsilon_{2, \, x} \sqrt{1 - \beta^2} \sin{\varphi_1} - 7  \epsilon_{1, \, x}  \epsilon_{2, \, y} \sqrt{1 - \beta^2} \sin{\varphi_1} - \cos{\varphi_2} (4  \epsilon_{1, \, x}  \epsilon_{2, \, x} + 7  \epsilon_{1, \, y}  \epsilon_{2, \, y} 
\\
\notag
&+& (4  \epsilon_{1, \, y}  \epsilon_{2, \, x} - 7  \epsilon_{1, \, x}  \epsilon_{2, \, y}) \sqrt{1 - \beta^2} \sin{\varphi_1}) - 
7  \epsilon_{1, \, y}  \epsilon_{2, \, x} \sqrt{1 - \beta^2} \sin{\varphi_2} + 4  \epsilon_{1, \, x}  \epsilon_{2, \, y} \sqrt{1 - \beta^2} \sin{\varphi_2} 
\\
\notag
&+& \cos{\varphi_1} (-4  \epsilon_{1, \, x}  \epsilon_{2, \, x} - 7  \epsilon_{1, \, y}  \epsilon_{2, \, y} + 2 (6  \epsilon_{1, \, x}  \epsilon_{2, \, x} + 5  \epsilon_{1, \, y}  \epsilon_{2, \, y}) \cos{\varphi_2} + 
       (7  \epsilon_{1, \, y}  \epsilon_{2, \, x} - 4  \epsilon_{1, \, x}  \epsilon_{2, \, y}) \sqrt{1 - \beta^2} \sin{\varphi_2}) 
\\
\notag
&+& 4 ( \epsilon_{1, \, x}  \epsilon_{2, \, x} -  \epsilon_{1, \, y}  \epsilon_{2, \, y}) (-1 + \sin{\varphi_1} \sin{\varphi_2})) + 
    \epsilon_{3, \, y}^* ((4  \epsilon_{1, \, y}  \epsilon_{2, \, x} - 7  \epsilon_{1, \, x}  \epsilon_{2, \, y}) \cos{\varphi_2} 
    \\
    \notag
    &+& \cos{\varphi_1} (-7  \epsilon_{1, \, y}  \epsilon_{2, \, x} + 4  \epsilon_{1, \, x}  \epsilon_{2, \, y} + 10 ( \epsilon_{1, \, y}  \epsilon_{2, \, x} +  \epsilon_{1, \, x}  \epsilon_{2, \, y}) \cos{\varphi_2}) - 7  \epsilon_{1, \, x}  \epsilon_{2, \, x} \sqrt{1 - \beta^2} \sin{\varphi_1} 
    \\
    \notag
    &-& 4  \epsilon_{1, \, y}  \epsilon_{2, \, y} \sqrt{1 - \beta^2} \sin{\varphi_1} - 7  \epsilon_{1, \, x}  \epsilon_{2, \, x} \sqrt{1 - \beta^2} \sin{\varphi_2} - 4  \epsilon_{1, \, y}  \epsilon_{2, \, y} \sqrt{1 - \beta^2} \sin{\varphi_2} 
    \\
    \notag
    &+& 7 ( \epsilon_{1, \, y}  \epsilon_{2, \, x} +  \epsilon_{1, \, x}  \epsilon_{2, \, y}) (-1 + \sin{\varphi_1} \sin{\varphi_2}) + 
     7  \epsilon_{1, \, x}  \epsilon_{2, \, x} \sqrt{1 - \beta^2} \sin{(\varphi_1 + \varphi_2)} 
     \\ 
     &+& 4  \epsilon_{1, \, y}  \epsilon_{2, \, y} \sqrt{1 - \beta^2} \sin{(\varphi_1 + \varphi_2)}))/8,    
\end{eqnarray}
\begin{eqnarray}
\notag
p_{12, \, z}^3 &=& -\beta^3 ( \epsilon_{3, \, x}^* ((7  \epsilon_{1, \, y}  \epsilon_{2, \, x} - 4  \epsilon_{1, \, x}  \epsilon_{2, \, y}) \cos{\varphi_1} + ( \epsilon_{1, \, y}  \epsilon_{2, \, x} +  \epsilon_{1, \, x}  \epsilon_{2, \, y}) (-10 + 7 \cos{(\varphi_1 - \varphi_2)}) 
\\
\notag
&+& (-4  \epsilon_{1, \, y}  \epsilon_{2, \, x} + 7  \epsilon_{1, \, x}  \epsilon_{2, \, y}) \cos{\varphi_2}) + 
\epsilon_{3, \, y}^* (-2 (5  \epsilon_{1, \, x}  \epsilon_{2, \, x} + 6  \epsilon_{1, \, y}  \epsilon_{2, \, y}) + (7  \epsilon_{1, \, x}  \epsilon_{2, \, x} 
\\
&+& 4  \epsilon_{1, \, y}  \epsilon_{2, \, y}) \cos{\varphi_1} + 4 (-( \epsilon_{1, \, x}  \epsilon_{2, \, x}) +  \epsilon_{1, \, y}  \epsilon_{2, \, y}) \cos{(\varphi_1 - \varphi_2)} + (7  \epsilon_{1, \, x}  \epsilon_{2, \, x} + 4  \epsilon_{1, \, y}  \epsilon_{2, \, y}) \cos{\varphi_2}))/8.
\end{eqnarray}
Similary, for $\mathbf{m}_{12}^3$ we have:
\begin{eqnarray}
\notag
m_{12, \, x}^3 &=& \beta^2 (-2  \epsilon_{3, \, x}^* (4 (-( \epsilon_{1, \, x}  \epsilon_{2, \, x}) +  \epsilon_{1, \, y}  \epsilon_{2, \, y}) \sqrt{1 - \beta^2} + 4  \epsilon_{1, \, y}  \epsilon_{2, \, x} \sin{\varphi_1} - 7  \epsilon_{1, \, x}  \epsilon_{2, \, y} \sin{\varphi_1} 
\\
\notag
&-& \cos{\varphi_2} ((4  \epsilon_{1, \, x}  \epsilon_{2, \, x} + 7  \epsilon_{1, \, y}  \epsilon_{2, \, y}) \sqrt{1 - \beta^2} + (4  \epsilon_{1, \, y}  \epsilon_{2, \, x} - 7  \epsilon_{1, \, x}  \epsilon_{2, \, y}) \sin{\varphi_1}) - 7  \epsilon_{1, \, y}  \epsilon_{2, \, x} \sin{\varphi_2} + 4  \epsilon_{1, \, x}  \epsilon_{2, \, y} \sin{\varphi_2} 
\\
\notag
&+& 4  \epsilon_{1, \, x}  \epsilon_{2, \, x} \sqrt{1 - \beta^2} \sin{\varphi_1} \sin{\varphi_2} - 4  \epsilon_{1, \, y}  \epsilon_{2, \, y} \sqrt{1 - \beta^2} \sin{\varphi_1} \sin{\varphi_2} + \cos{\varphi_1} (-((4  \epsilon_{1, \, x}  \epsilon_{2, \, x} + 7  \epsilon_{1, \, y}  \epsilon_{2, \, y}) \sqrt{1 - \beta^2}) 
\\
\notag
&+& 2 (6  \epsilon_{1, \, x}  \epsilon_{2, \, x} + 5  \epsilon_{1, \, y}  \epsilon_{2, \, y}) \sqrt{1 - \beta^2} \cos{\varphi_2} + 
(7  \epsilon_{1, \, y}  \epsilon_{2, \, x} - 4  \epsilon_{1, \, x}  \epsilon_{2, \, y}) \sin{\varphi_2})) 
\\
\notag
&+& \epsilon_{3, \, y}^* (14 ( \epsilon_{1, \, y}  \epsilon_{2, \, x} +  \epsilon_{1, \, x}  \epsilon_{2, \, y}) \sqrt{1 - \beta^2} + 2 (7  \epsilon_{1, \, y}  \epsilon_{2, \, x} - 4  \epsilon_{1, \, x}  \epsilon_{2, \, y}) \sqrt{1 - \beta^2} \cos{\varphi_1}
\\
\notag
&-& 17 ( \epsilon_{1, \, y}  \epsilon_{2, \, x} +  \epsilon_{1, \, x}  \epsilon_{2, \, y}) \sqrt{1 - \beta^2} \cos{(\varphi_1 - \varphi_2)} - 8  \epsilon_{1, \, y}  \epsilon_{2, \, x} \sqrt{1 - \beta^2} \cos{\varphi_2} + 14  \epsilon_{1, \, x}  \epsilon_{2, \, y} \sqrt{1 - \beta^2} \cos{\varphi_2} 
\\
\notag
&-& 3  \epsilon_{1, \, y}  \epsilon_{2, \, x} \sqrt{1 - \beta^2} \cos{(\varphi_1 + \varphi_2)} - 3  \epsilon_{1, \, x}  \epsilon_{2, \, y} \sqrt{1 - \beta^2} \cos{(\varphi_1 + \varphi_2)} 
\\
&+& 2 (7  \epsilon_{1, \, x}  \epsilon_{2, \, x} + 4  \epsilon_{1, \, y}  \epsilon_{2, \, y}) (\sin{\varphi_1} + \sin{\varphi_2} - \sin{(\varphi_1 + \varphi_2)})))/16,
\\
\notag
\\
\notag
m_{12, \, y}^3 &=& \beta^2 ( \epsilon_{3, \, y}^* (7  \epsilon_{1, \, y}  \epsilon_{2, \, x} \sqrt{1 - \beta^2} \sin{\varphi_1} - 4  \epsilon_{1, \, x}  \epsilon_{2, \, y} \sqrt{1 - \beta^2} \sin{\varphi_1} - \cos{\varphi_2} (7  \epsilon_{1, \, x}  \epsilon_{2, \, x} + 4  \epsilon_{1, \, y}  \epsilon_{2, \, y}
\\
\notag
&+& (7  \epsilon_{1, \, y}  \epsilon_{2, \, x} - 4  \epsilon_{1, \, x}  \epsilon_{2, \, y}) \sqrt{1 - \beta^2} \sin{\varphi_1}) - 
     4  \epsilon_{1, \, y}  \epsilon_{2, \, x} \sqrt{1 - \beta^2} \sin{\varphi_2} + 7  \epsilon_{1, \, x}  \epsilon_{2, \, y} \sqrt{1 - \beta^2} \sin{\varphi_2} 
     \\
     \notag
     &+& \cos{\varphi_1} (-7  \epsilon_{1, \, x}  \epsilon_{2, \, x} - 4  \epsilon_{1, \, y}  \epsilon_{2, \, y} + 2 (5  \epsilon_{1, \, x}  \epsilon_{2, \, x} + 6  \epsilon_{1, \, y}  \epsilon_{2, \, y}) \cos{\varphi_2} + 
       (4  \epsilon_{1, \, y}  \epsilon_{2, \, x} - 7  \epsilon_{1, \, x}  \epsilon_{2, \, y}) \sqrt{1 - \beta^2} \sin{\varphi_2}) 
       \\
       \notag
       &-& 4 ( \epsilon_{1, \, x}  \epsilon_{2, \, x} -  \epsilon_{1, \, y}  \epsilon_{2, \, y}) (-1 + \sin{\varphi_1} \sin{\varphi_2})) + 
    \epsilon_{3, \, x}^* ((-7  \epsilon_{1, \, y}  \epsilon_{2, \, x} + 4  \epsilon_{1, \, x}  \epsilon_{2, \, y}) \cos{\varphi_2} 
    \\
    \notag
    &+& \cos{\varphi_1} (4  \epsilon_{1, \, y}  \epsilon_{2, \, x} - 7  \epsilon_{1, \, x}  \epsilon_{2, \, y} + 10 ( \epsilon_{1, \, y}  \epsilon_{2, \, x} +  \epsilon_{1, \, x}  \epsilon_{2, \, y}) \cos{\varphi_2}) + 4  \epsilon_{1, \, x}  \epsilon_{2, \, x} \sqrt{1 - \beta^2} \sin{\varphi_1} 
    \\
    \notag
    &+& 7  \epsilon_{1, \, y}  \epsilon_{2, \, y} \sqrt{1 - \beta^2} \sin{\varphi_1} + 4  \epsilon_{1, \, x}  \epsilon_{2, \, x} \sqrt{1 - \beta^2} \sin{\varphi_2} + 7  \epsilon_{1, \, y}  \epsilon_{2, \, y} \sqrt{1 - \beta^2} \sin{\varphi_2} 
    \\ 
    \notag
    &+& 7 ( \epsilon_{1, \, y}  \epsilon_{2, \, x} +  \epsilon_{1, \, x}  \epsilon_{2, \, y}) (-1 + \sin{\varphi_1} \sin{\varphi_2}) - 
     4  \epsilon_{1, \, x}  \epsilon_{2, \, x} \sqrt{1 - \beta^2} \sin{(\varphi_1 + \varphi_2)} 
     \\
     &-& 7  \epsilon_{1, \, y}  \epsilon_{2, \, y} \sqrt{1 - \beta^2} \sin{(\varphi_1 + \varphi_2)}))/8, 
\\
\notag
\\
\notag
m_{12, \, z}^3 &=& \beta^3 ( \epsilon_{3, \, y}^* ((-4  \epsilon_{1, \, y}  \epsilon_{2, \, x} + 7  \epsilon_{1, \, x}  \epsilon_{2, \, y}) \cos{\varphi_1} + ( \epsilon_{1, \, y}  \epsilon_{2, \, x} +  \epsilon_{1, \, x}  \epsilon_{2, \, y}) (-10 + 7 \cos{(\varphi_1 - \varphi_2)}) 
\\
\notag
&+& (7  \epsilon_{1, \, y}  \epsilon_{2, \, x} - 4  \epsilon_{1, \, x}  \epsilon_{2, \, y}) \cos{\varphi_2}) + 
    \epsilon_{3, \, x}^* (-2 (6  \epsilon_{1, \, x}  \epsilon_{2, \, x} + 5  \epsilon_{1, \, y}  \epsilon_{2, \, y}) 
    + (4  \epsilon_{1, \, x}  \epsilon_{2, \, x} + 7  \epsilon_{1, \, y}  \epsilon_{2, \, y}) \cos{\varphi_1} 
    \\
    &+& 4 ( \epsilon_{1, \, x}  \epsilon_{2, \, x} -  \epsilon_{1, \, y}  \epsilon_{2, \, y}) \cos{(\varphi_1 - \varphi_2)} + (4  \epsilon_{1, \, x}  \epsilon_{2, \, x} + 7  \epsilon_{1, \, y}  \epsilon_{2, \, y}) \cos{\varphi_2}))/8.   
\end{eqnarray}

Components of symmetric matrix~$M$ in terms of the parameters~$a_l$,~$\Delta_l$ and geometry parameters of the scheme read: 
\begin{eqnarray}
\notag
M_{11} &=& 2 (\nu_1^2 a_1^2 + \nu_2^2 a_2^2 + \nu_3^2 a_3^2),
\\
\notag
M_{22} &=& 2(\nu_1^2 \Delta_1^2 + \nu_2^2 \Delta_2^2 + \nu_3^2 \Delta_3^2 - 
\beta^2 (\nu_1^2 (\Delta_1^2 - a_1^2) \cos^2 \varphi_1 + 
\nu_2^2 (\Delta_2^2 - a_2^2) \cos^2 \varphi_2 +
\nu_3^2 (\Delta_3^2 - a_3^2))),
\\
\notag
M_{33} &=& 2(\nu_1^2 \Delta_1^2 + \nu_2^2 \Delta_2^2 + \nu_3^2 \Delta_3^2
- \beta^2 (\nu_1^2 (\Delta_1^2 - a_1^2) \sin^2 \varphi_1 + 
\nu_2^2 (\Delta_2^2 - a_2^2) \sin^2 \varphi_2)),
\\
\notag
M_{44} &=& 2 ( \nu_1^2 a_1^2 + \nu_2^2 a_2^2 + \nu_3^2 a_3^2 + \beta^2 (\nu_1^2 (\Delta_1^2 - a_1^2) + \nu_2^2 (\Delta_2^2 - a_2^2) + \nu_3^2 (\Delta_3^2 - a_3^2) )),
\\
\notag
M_{12} &=& -2 \beta (\nu_1^2 a_1^2 \cos \varphi_1 + \nu_2^2 a_2^2 \cos \varphi_2 + \nu_3^2 a_3^2),
\\
\notag
M_{13} &=& -2 \beta (\nu_1^2 a_1^2 \sin \varphi_1 + \nu_2^2 a_2^2 \sin \varphi_2),
\\
\notag
M_{14} &=& -2 \sqrt{1 - \beta^2} (\nu_1^2 a_1^2 + \nu_2^2 a_2^2 + \nu_3^2 a_3^2),
\\
\notag
M_{23} &=&  -\beta^2 ( \nu_1^2 (\Delta_1^2 - a_1^2) \sin(2  \varphi_1) + \nu_2^2 ( \Delta_2^2 - a_2^2) \sin(2  \varphi_2)),
\\
\notag
M_{24} &=& -2 \beta \sqrt{1 - \beta^2} ( \nu_1^2  (\Delta_1^2 - a_1^2 ) \cos \varphi_1 + \nu_2^2 ( \Delta_2^2 - a_2^2) \cos \varphi_2 + \nu_3^2 ( \Delta_3^2 - a_3^2)),
\\
M_{34} &=& -2 \beta \sqrt{1 - \beta^2} ( \nu_1^2 (\Delta_1^2 - a_1^2) \sin \varphi_1 + \nu_2^2 (\Delta_2^2 - a_2^2) \sin \varphi_2).
\end{eqnarray}

Matrix~$M_B$ in terms of the geometry parameters of the scheme reads: 
\begin{equation}
M_B =
\left(
\begin{matrix}
1 & 0 & 0 \\
-\beta \cos \varphi_s & -\sqrt{1 - \beta^2} \nu_s \cos \varphi_s & \beta \nu_s \sin \varphi_s \\
-\beta \sin \varphi_s & -\sqrt{1 - \beta^2} \nu_s \sin \varphi_s & -\beta \nu_s \cos \varphi_s  \\
-\sqrt{1- \beta^2} & \beta \nu_s & 0
\end{matrix}
\right).
\end{equation}
\end{widetext}

\providecommand{\noopsort}[1]{}\providecommand{\singleletter}[1]{#1}%

\end{document}